\documentclass[useAMS,usenatbib]{mn2e}
\usepackage[]{natbib}
\usepackage{amsmath}
\usepackage{amssymb}
 \usepackage{graphicx}
 \usepackage{indentfirst}
 \usepackage{pdflscape}
 \usepackage{placeins}
 \usepackage{nicefrac}
 \usepackage{afterpage}
 \usepackage{rotating}
 \newcommand{\kms}{${\rm km\,s}^{-1}$}
 \newcommand{\etal}{et al.}
\newcommand{\lya}{\mbox{${\rm Ly}\alpha$}}
\newcommand{\lyb}{\mbox{${\rm Ly}\beta$}}
\newcommand{\ovi}{O\,VI}

\title[]{Probing the IGM-galaxy connection at $z<0.5$ II. New insights into the galaxy environments of O\,VI absorbers in PKS 0405$-$123}
\author[]{Sean D. Johnson$^{1}$\thanks{E-mail:
seanjohnson@uchicago.edu}, Hsiao-Wen Chen$^{1}$\thanks{E-mail:
hchen@oddjob.uchicago.edu}, John S. Mulchaey$^{2}$ \\
$^{1}$Department of Astronomy \& Astrophysics and Kavli Institute for Cosmological Physics, The University of Chicago, Chicago, IL 60637, USA\\
$^{2}$The Observatories of the Carnegie Institute of Washington, 813 Santa Barbara Street, Pasadena, CA 91101, USA}
\begin{document}

\date{\today}

\pagerange{\pageref{firstpage}--\pageref{lastpage}} \pubyear{2013}

\maketitle

\label{firstpage}

\begin{abstract}

  We present new absorption-line analysis and new galaxy survey data
  obtained for the field around PKS\,0405$-$123 at $z_{\rm QSO}=0.57$.
  Combining previously known O\,VI absorbers with new identifications
  in the higher $S/N$ UV spectra obtained with the Cosmic Origins
  Spectrograph, we have established a sample of seven O\,VI absorbers
  and 12 individual components at $z=0.0918-0.495$ along the sightline
  toward PKS\,0405$-$123.  We complement the available UV absorption
  spectra with galaxy survey data that reach $100$\% completeness at
  projected distances $\rho<200$ kpc of the quasar sightline for galaxies
  as faint as $0.1\,L_*$ ($0.2\,L_*$) out to redshifts of
  $z \approx 0.35$ ($z \approx 0.5$).  The high level of completeness
  achieved at faint magnitudes by our survey reveals that O\,VI
  absorbers are closely associated with gas-rich environments
  containing at least one low-mass, emission-line galaxy.  An
  intriguing exception is a strong O\,VI system at $z \approx 0.183$
  that does not have a galaxy found at $\rho<4$ Mpc, and our survey
  rules out the presence of any galaxies of $L>0.04\,L_*$ at
  $\rho<250$ kpc and any galaxies of $L>0.3\,L_*$ at $\rho<1$
  Mpc.  We further examine the galactic environments of O\,VI
  absorbers and those ``\lya-only'' absorbers with neutral hydrogen
  column density $\log\,N({\rm H\,I})>13.6$ and no detectable O\,VI
  absorption features.  The \lya-only absorbers serve as a control
  sample in seeking the discriminating galactic features that result
  in the observed O\,VI absorbing gas at large galactic radii.  We
  find a clear distinction in the radial profiles of mean galaxy
  surface brightness around different absorbers.
  Specifically, O\,VI absorbers are found to reside in regions of higher
  mean surface brightness at $\rho\lesssim 500$ kpc 
  ($\Delta \mu_R \approx +5 \, {\rm mag \, Mpc^{-2}}$ relative to the
   background at $\rho>500$ kpc), while only a mild
  increase in galaxy surface brightness is seen at small $\rho$ around
  \lya-only absorbers ($\Delta \mu_R \approx +2\, {\rm mag \, Mpc^{-2}}$).  
  The additional insights gained from our deep galaxy survey 
  demonstrates the need to probe the galaxy populations
  to low luminosities in order to better understand the nature of the
  absorbing systems.

\end{abstract}


\section{Introduction}

  \begin{figure*}
	\centering
	\includegraphics[angle=270,scale=0.65]{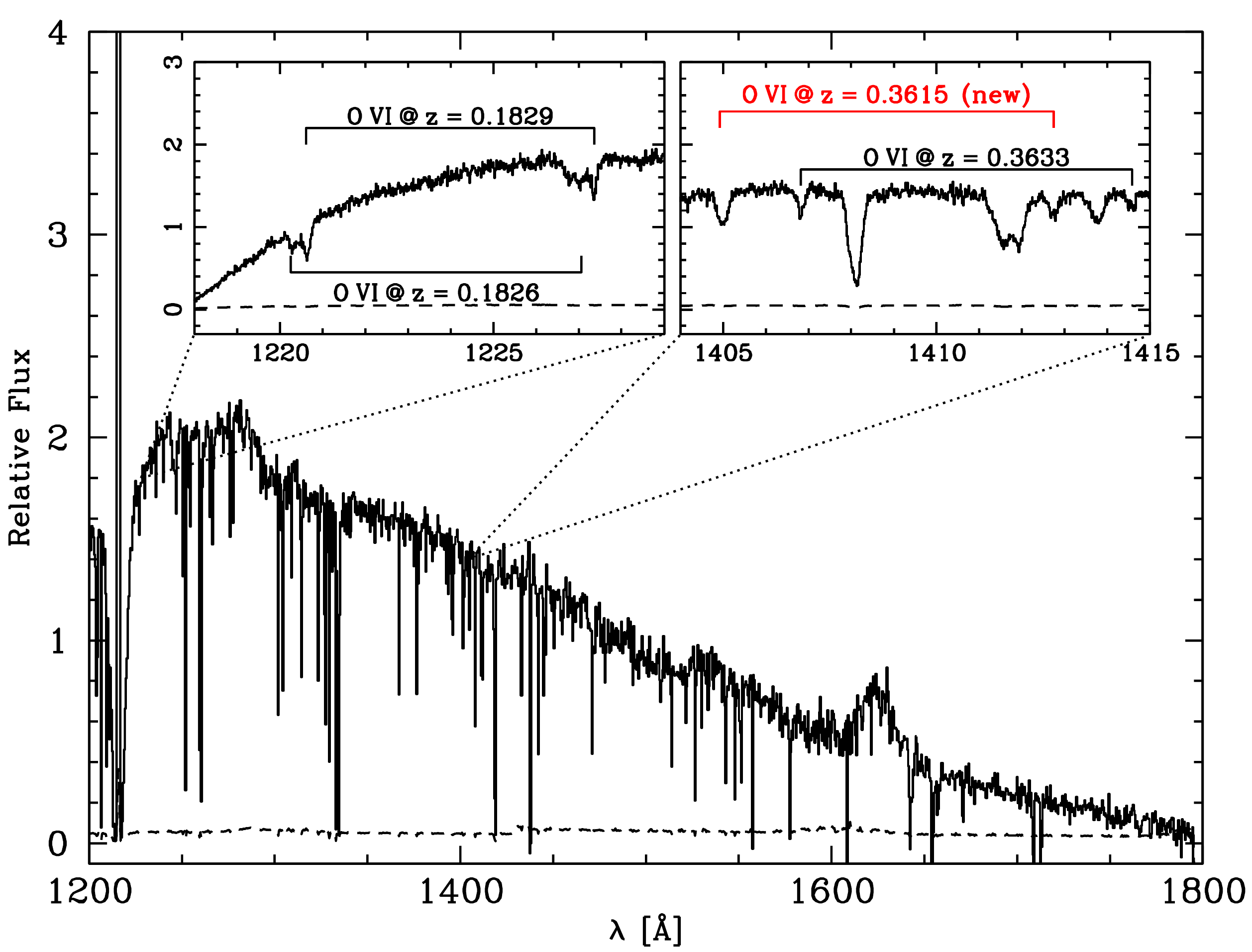}
	\caption{Reduced and combined COS spectrum (solid line) of
          PKS\,0405$-$123 with corresponding 1-$\sigma$ errors (dashed
          line). The inset plots highlight detections enabled by
          the improved $S/N$ of the quasar spectrum. {\it Left}: O\,VI
          absorption at $z=0.1829$ and 0.1826. {\it Right}:
          O\,VI absorption at $z=0.3633$ and 0.3615 (new). }
	\label{figure:COS}
\end{figure*}

Ultraviolet absorption lines in the spectra of background sources
represent the most sensitive available means of observing the diffuse
gas that permeates the universe.  The O\,VI $\lambda\lambda$1031, 1037
doublet in particular has received attention as a tracer of: the
warm-hot phase of the intergalactic medium \citep[IGM;
e.g.][]{Cen:1999}, the galaxy outflows thought to be responsible for
the chemical enrichment of the IGM \citep[e.g.][]{Oppenheimer:2006},
and the intra-group medium \citep{Mulchaey:1996}.  While independent
surveys of O\,VI absorbers in the spectra of distant quasars have
uncovered a large number of these systems supporting the notion of
O\,VI doublets being a sensitive tracer of warm-hot gas, the reported
number density of O\,VI absorbers from different surveys 
shows a scatter much beyond the individual measurement errors 
\citep[e.g.][]{Burles:1996, Tripp:2000, TrippSavage:2000, Prochaska:2004, Richter:2004, Tripp:2008, Thom:2008a, Danforth:2008, Tilton:2012}. 
At the same time, cosmological simulations incorporating momentum-driven
winds have been able to reproduce the observed O\,VI absorption column
density distribution function, but ambiguity remains in attributing the
majority of low-redshift O\,VI absorbers to either cool, photoionized gas
\citep[e.g.][]{Kang:2005, Oppenheimer:2009, Oppenheimer:2012} or the
warm-hot phase of the IGM \citep[cf.][]{Smith:2011, Tepper:2011,
Cen:2012, Stinson:2012}.

Key insights into the physical origin of O\,VI absorbers can be gained
from a detailed examination of their galactic environments.
Observations designed to constrain the properties of gaseous halos of
known galaxies have shown that emission-line galaxies exhibit near
unity O\,VI covering fractions ($\kappa_{\rm O\,VI}$) at projected
distances $\rho \lesssim 150$ kpc and $\kappa_{\rm O\,VI}\approx 64$\%
at $\rho < 350$ kpc, while absorption-line galaxies exhibit
$\kappa_{\rm O\,VI}\lesssim 30$\% on similar scales \citep{Chen:2009,
  Tumlinson:2011}.  The large incidence of O\,VI absorbers around
star-forming galaxies may be explained by a causal connection between
star formation and the production of O\,VI absorbing gas, but the
non-negligible covering fraction of such gas around an evolved galaxy
population becomes difficult to explain under the same scenario. 

While surveys of galaxies associated with known O\,VI absorbers have
revealed a correlation between the presence of star-forming galaxies
and O\,VI absorbers at modest projected separations $\rho\lesssim 350$
kpc \citep{Stocke:2006, Prochaska:2006, Chen:2009, Mulchaey:2009, Wakker:2009, Stocke:2013}, the galaxy survey data are not sufficiently deep and
complete for a detailed examination of the galactic environment
immediate to the absorbers.  Dedicated surveys around UV bright
quasar sightlines are typically limited to bright galaxies with $R$-band
magnitudes brighter than $AB(R)\approx 19.5$ limiting sensitivity to
$L\approx 0.1\,L_*$ galaxies at $z\lesssim0.1$
\citep[e.g.][]{Prochaska:2011survey}.  Although the study of
\cite{Wakker:2009} includes galaxies fainter than $L=0.1\,L_*$ at
$z<0.02$, the incompleteness of their galaxy catalog is unknown.
Survey incompleteness complicates the interpretation of the
galaxy--absorber studies \citep{Stocke:2006, Stocke:2013}.  To date, a
high completeness level ($> 95$\%) for galaxies of $R\lesssim 23$ and
$\Delta \theta \lesssim 2'$ from the quasar sightline has been reached in only one quasar field
(HE\,0226$-$4110 in Chen \& Mulchaey 2009).  To improve the
statistics, our group is continuing the effort to collect
high-completeness galaxy survey data in multiple quasar fields.

In this paper, we present new galaxy survey data obtained for the
field around PKS\,0405$-$123 at $z_{\rm QSO}=0.57$.  Our galaxy survey
in this field has reached $100$\% completeness within $\rho=200$ kpc
of the quasar sightline for galaxies as faint as $0.1\,L_*$ ($0.2\,L_*$)
out to redshifts of $z\approx0.35$ ($z\approx 0.5$).  PKS\,0405$-$123
is among the brightest quasars on the sky, for which high-quality UV
echelle data have been obtained using the Space Telescope Imaging
Spectrograph \citep[STIS;][]{Woodgate:1998} and extremely high-quality
UV spectra have been obtained using the new Cosmic Origins
Spectrograph \citep[COS;][]{Green:2012} on board the {\it Hubble Space
  Telescope} ({\it HST}).  The intermediate redshift of the quasar provides a
long redshift path length for probing intervening absorption systems.
Previous systematic searches in the STIS and 
Far Ultraviolet Spectroscopic Explorer \citep[FUSE;][]{Moos:2000} spectra have uncovered six
O\,VI absorption systems at $z_{\rm O\,VI}=0.09-0.5$
\citep{Prochaska:2004, Howk:2009} and 11 additional strong \lya\
absorbers of neutral hydrogen column density $\log\,N({\rm H\,I})\ge
13.6$ at $z_{\rm H\,I}=0.03-0.5$ \citep[e.g.][]{Williger:2006,
  Lehner:2007}.  Recent targeted searches in the new COS spectra have
further uncovered a Ne\,VIII absorber associated with the strong O\,VI
absorber at $z=0.495$ \citep{Narayanan:2011} and a new component
associated with the previously known O\,VI absorber at $z=0.167$
\citep{Savage:2010}.  

To complement the available highly complete galaxy survey data, we
have carried out a new systematic search of absorption
features in the new COS spectra.  Here we discuss new insights into
the origin of O\,VI absorbing gas that we have learned from combining
the improved absorption-line measurements and highly complete galaxy
survey data.
The paper proceeds as follows: In Section \ref{section:data} we
present the full archival COS spectrum of PKS 0405$-$123 and new galaxy
redshifts in our survey.  In Section \ref{section:new_abs} we review
measurements of the absorbers enabled by the COS spectrum including (1) a
tentative detection of an O\,VI absorber at $z=0.2977$, (2) a new
O\,VI absorber at $z=0.3615$, and (3) N\,V associated with a known
O\,VI absorber at $z=0.3633$.  In Section \ref{section:environments} we
discuss the galaxy environments of all seven O\,VI absorbers along this 
sightline and compare them to the galaxy environments of strong \lya\ 
absorbers with no detected O\,VI.  Finally, in Section \ref{section:discussion},
we briefly discuss the implications of our findings.

Throughout the paper, we adopt a $\Lambda$ cosmology with
$\Omega_{\rm m}=0.3$, $\Omega_\Lambda=0.7$, and $H_0 = 70$ \kms$\,{\rm Mpc}^{-1}$.  We also
adopt a non-evolving, rest-frame absolute $R$-band magnitude (AB) of
$M_{R*}= -21.17$ for $L_*$ galaxies based on \cite{Blanton:2003}.
Unless otherwise stated, we perform $k$-corrections using the Scd
galaxy template from \cite{Coleman:1980}.

\begin{figure}
	\includegraphics[angle=0,scale=0.43]{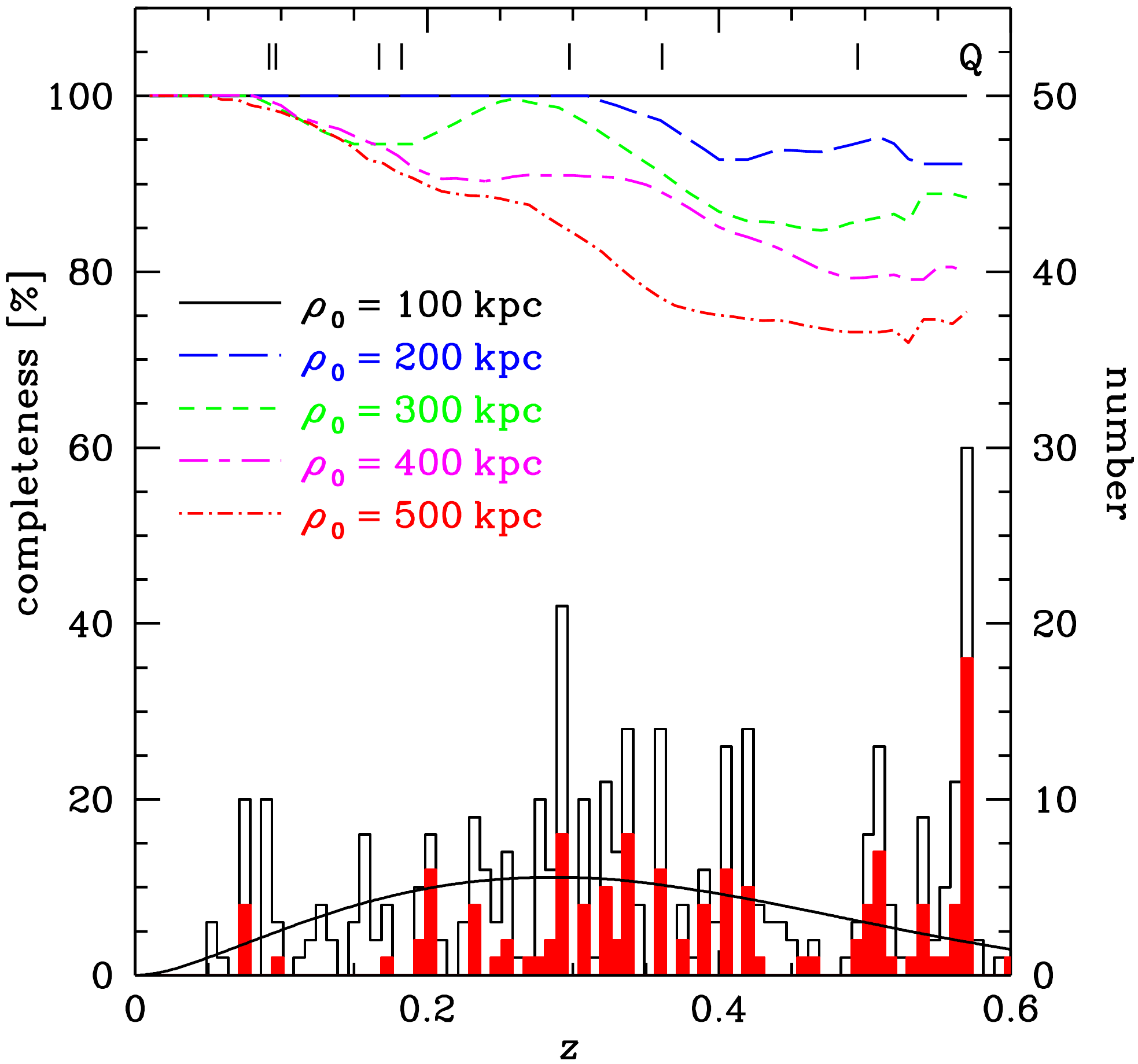}
	\caption{Summary of our galaxy survey results.  The curves at
          the top show the estimated survey completeness for $L >
          0.1\,L_*$ galaxies as a function of redshift at projected
          distances of $\rho<100$, $200$, $300$, $400$, and $500$ kpc
          of the quasar sightline. The bottom histograms show the
          redshift distributions of all galaxies (black) and
          absorption-line dominated galaxies (red, solid) in the final
          combined spectroscopic catalog.  For comparison, we show the
          expected redshift distribution based on a non-evolving
          $R$-band luminosity function adapted from
          \protect\cite{Blanton:2003}, taking into account our survey
          incompleteness as a function of galaxy luminosity and
          redshift (solid black line). Spikes in the histogram that
          deviate significantly from expectations are due to
          large-scale galaxy overdensities in the quasar field. The
          redshifts of O\,VI absorbers (vertical ticks) and the quasar
          (Q) are shown along the top of the figure.}
	\label{figure:completeness}
\end{figure}

\section{Data}
\label{section:data}
PKS\,0405$-$123 is a well-studied sightline with available imaging and
UV spectroscopic data in the FUSE and {\it HST} archives \citep{Chen:2000,
  Prochaska:2004, Williger:2006, Lehner:2007}.  Different galaxy
surveys have been carried out in this field identifying galaxies
associated with absorption-line systems uncovered in the UV spectra
\citep{Spinrad:1993, Ellingson:1994, Chen:2005, Prochaska:2006,
  Chen:2009}.  Recently, PKS\,0405$-$123 was targeted for {\it HST}/COS UV
spectroscopy at significantly higher $S/N$ than the archival
STIS spectra.  This new data set led to the discovery of an
interesting O\,VI absorber with no detectable H\,I at $\Delta v
\approx-300$ \kms \ from a previously detected O\,VI absorber in a
partial Lyman-limit system at $z = 0.1671$ \citep{Savage:2010}.  We
have analyzed the full COS spectrum of PKS\,0405$-$123 (Figure
\ref{figure:COS}) together with new and existing galaxy survey data
for a comprehensive study of the galactic environments of absorbing
clouds uncovered along this quasar sightline.

\begin{figure*}
	\centering
	\includegraphics[angle=0,scale=0.65]{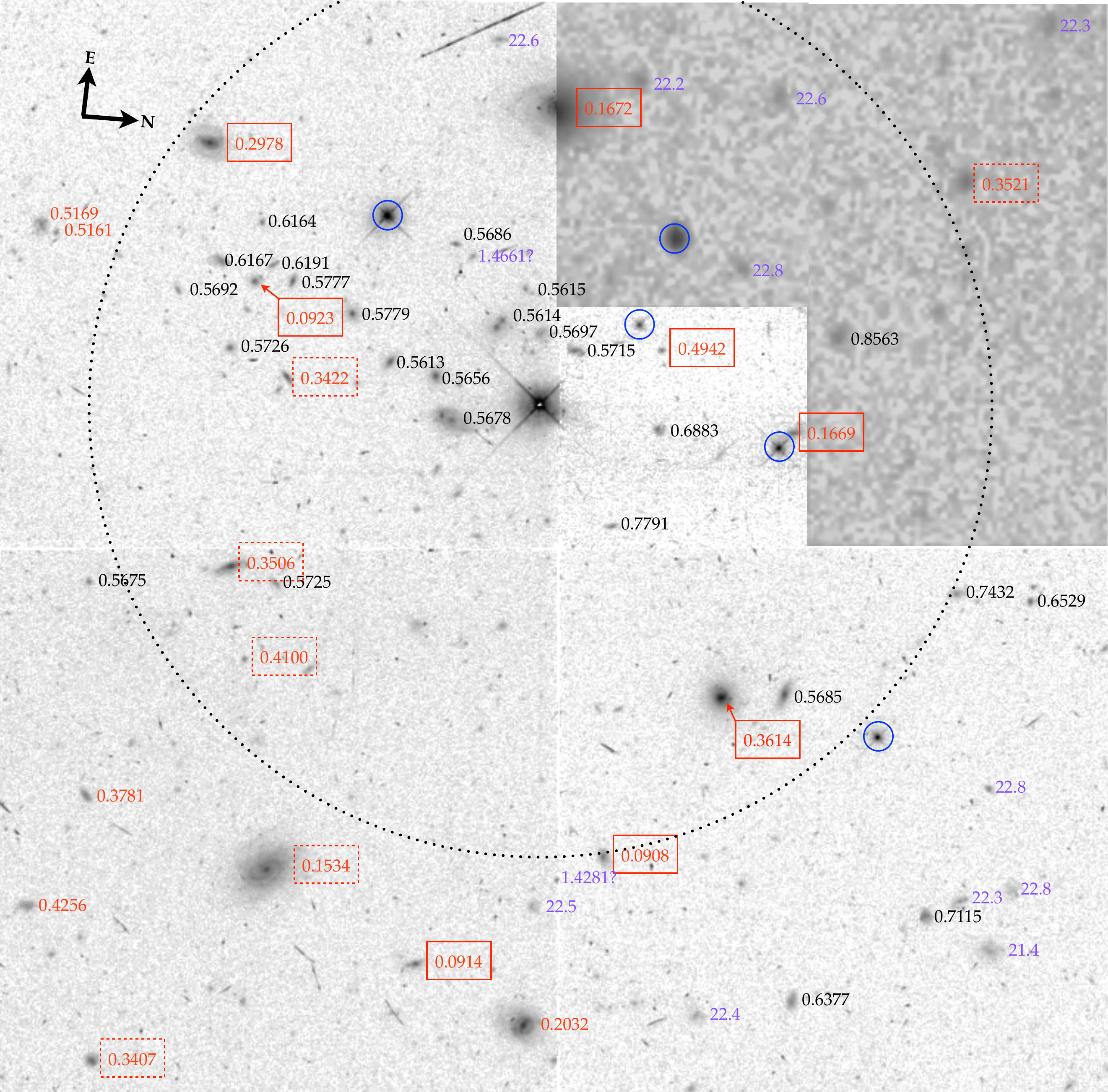}
	\caption{{\it HST}/Wide Field Planetary Camera 2 image of PKS\,0405$-$123 augmented with an
          image from the DuPont telescope
          \protect\citep{Prochaska:2006} where {\it HST} imagery is
          unavailable.  The orientation of the field is indicated by
          the N--E arrows that appear in the upper-left corner.
          Redshifts are shown to the right of the corresponding galaxy
          except in crowded areas where an arrow is used to indicate
          association.  Galaxies foreground to the quasar are shown in
          red, while background galaxies ($z \gtrsim
          z_{\mathrm{QSO}}$) are shown in black.  Galaxies associated
          with O\,VI absorbers (with velocity offsets $|\Delta v| <
          300$ \kms) are marked by solid red boxes, while galaxies
          associated with Ly$\alpha$ absorbers and no detectable O\,VI
          are marked by dotted red boxes.  Stars are labelled with
          blue circles and galaxies without secure redshifts are
          marked in purple either by redshift (based on a single
          emission line) or $R$-band magnitude if no redshift is
          available.  A dotted circle centered on the quasar with $1'$
          radius is shown to provide scale.  In addition to the many
          foreground galaxies spectroscopically identified near the quasar
          line-of-sight, there exists a clear overdensity of galaxies
          at the redshift of the quasar.  }
	\label{figure:hst}
\end{figure*}

\begin{table*}
\centering
\caption{Journal of spectroscopic observations}
\label{table:observations}
\begin{tabular}{lccccc}
 \hline
 \hline
  &  & FWHM  &  & Exposure time &  \\
 Telescope & Instrument/setup & (\AA) & No. of targets & (ks) & Date \\
  \hline
 Magellan Baade & IMACS/f2/200l, $1''$ slitlets & $9$ & $17$ & $3.6$ & Nov. $2012$ \\
   &  &  &  $12$ & $3.0$ & Feb. $2011$ \\
  &  &  &  $150$ & $5.4$ & Nov. $2012$ \\
   &  &  &  $140$ & $5.4$ & Feb. $2011$ \\
  Magellan Clay  & LDSS3/VPH-all $1''$ slitlets & $10$ & $24$ & $7.2$ & Feb. $2011$\\
  APO $3.5$-m &  DIS/B$400$ \& R$300$, $1.5''$ long slit & $5$ &  $23$ & varies & Oct.$-$Dec. $2012$\\
 \hline
\end{tabular}
\end{table*}

\subsection{COS spectroscopy}
PKS\,0405$-$123 was targeted for COS observations by two separate {\it HST}
programmes (PI: Keith Noll, PID$=$11508 and PI: James Green,
PID$=$11541). The two programs together acquired 17 exposures,
totaling 22.2 ks with the G130M grating (covering the spectral range
$1150-1450$ \AA \ and with full width at half-maximum spectral
resolution of ${\rm FWHM}=16$ \kms), and 4 exposures, totaling 11.1 ks
with the G160M grating ($1400-1800$ \AA, ${\rm FWHM}=16$ \kms). The
exposures were acquired at different central wavelengths in order to
provide contiguous wavelength coverage despite the gap between COS
detectors.  We retrieved the 1-D individual calibrated spectra from
the {\it HST} archive and combined them using a custom suite of software
(see \citealt{Yoon:2012} for details).  Specifically, individual
spectra from the two detector segments were aligned and co-added using
a common Milky Way absorption line as a reference (e.g.,
Si\,III\,1206) and then combined into a single
one-dimensional spectrum.  These co-addition routines work with photon
counts rather than flux calibrated data to allow for an accurate error
estimate in the low-count regime \citep[e.g.][]{Gehrels:1986}.  A
well-known issue with far-ultraviolet (FUV) spectra obtained using COS is the presence
of fixed-pattern noise due to the grid-wire in the COS FUV
detectors. Such pattern noise can be reduced by the use of multiple
FP-POS settings (see the COS Instrument Handbook).  As described
above, PKS\,0405$-$123 was observed by two different programs and in
many exposures of different FP-POS settings.  The effect of such
fixed-pattern noise is therefore minimal in the final stack.  To ensure
consistency with previous results, we set a common zero-point by
aligning the co-added COS spectrum with the archival FUSE and STIS
spectra prior to combination of the G130M and G160M data.  This
alignment led to minor ($\ll$FWHM) higher-order corrections to the COS
wavelength calibration. By comparing the COS spectrum with the
archival STIS spectrum, we estimate that errors in the wavelength
calibration limit the accuracy of line-centroids to $\approx 4$ \kms.

The resulting high resolution spectrum
(see Figure \ref{figure:COS}) enabled the new detections of O\,VI systems in
addition to a number of other lines and significantly stronger upper
limits on key ions. The new detections are discussed in detail in
Section \ref{section:new_abs}.

\subsection{Galaxy Survey}

Previous spectroscopic surveys of galaxies in the PKS\,0405$-$123
field reached $\approx 80$\% completeness within a $3'$ radius of the
quasar sightline for galaxies brighter than $R=21$ and $\approx 50$\% for
galaxies between $R=21$ and $22$ (e.g. \citealt{Chen:2009}). These
surveys identified galaxies associated with half of the known
O\,VI absorbers along this quasar sightline; however, a detailed study of the
galaxy environments that includes sub-$L_*$ galaxies requires a higher
survey completeness at fainter magnitudes.  To improve the survey
depth, we acquired new galaxy spectra with the Magellan telescopes
using the Inamori-Magellan Areal Camera \& Spectrograph 
\citep[IMACS;][]{Dressler:2011} and the 
Low Dispersion Survey Spectrograph 3 (LDSS3)
(see Table \ref{table:observations} for a summary of the observations). In
addition, we confirmed a number of photometric stars using the Dual Imaging spectrograph (DIS)
 on the Apache Point 3.5-m.  The IMACS and LDSS3 data
were reduced using the Carnegie Observatories System for MultiObject Spectroscopy
(COSMOS\footnote{http://code.obs.carnegiescience.edu/cosmos})
as described in \cite{Chen:2009}. The APO DIS spectra were reduced
using a slightly modified version of the Low-REDUX pipeline written by
J. Hennawi, S. Burles, and
J. X. Prochaska\footnote{http://www.ucolick.org/$\sim$xavier/LowRedux/}.
Galaxy redshifts were determined both by cross-correlation with 
the Sloan Digital Sky Survey \citep[SDSS; ][]{York:2000}
templates and by fitting of galaxy eigenspectra as in
\cite{Chen:2009}.  In nearly all cases, the two independently
determined redshifts were in good agreement ($|\Delta z| \leq
0.0003$), but in the small number of cases for which they were not,
SDJ and HWC determined the best redshift by refitting and visual
inspection.  All assigned redshifts were visually inspected to
determine their reliability and galaxy spectra were further classified
as either absorption-line dominated or emission-line dominated. The
final object catalog is presented in Table \ref{table:galaxies}.

\begin{table*}
	\centering
	\caption{The photometric and spectroscopic catalog of galaxies around PKS 0405$-$123$^\mathrm{a}$. The full table is available on the journal webpage.}
	\label{table:galaxies}
	\begin{tabular}{ccccccccccc}
	\hline
	\hline
	 & RA & Dec & $\triangle\theta$ & $\rho$$^\mathrm{b}$ & $R$ &  &  &  &  &  \\
	ID & (J$2000$) & (J$2000$) & ($''$) & (kpc) & (mag) & $z_\mathrm{spec}$ & Quality$^\mathrm{c}$ & Object type$^\mathrm{d}$ & Galaxy class$^\mathrm{e}$ & $L/L_*^\mathrm{b,f}$ \\
	\hline
	$80001^{n}$ & 04:07:48.9 & -12:11:33 & $7.8$ & $35$ & $21.65 \pm 0.00$ & $0.5715$   & A & G & E & $0.95$  \\
	$90001^{}$ & 04:07:49.1 & -12:11:38 & $9.9$ & $45$ & $22.10 \pm 9.99$ & $0.5697$     & A & G & E & $0.62$  \\
	$80003^{n}$ & 04:07:49.1 & -12:11:43 & $11.7$ & $53$ & $21.10 \pm 0.00$ & $0.5709$ & A & G & A & $1.57$  \\
	$80004^{n}$ & 04:07:48.2 & -12:11:49 & $12.8$ & $58$ & $19.97 \pm 0.00$ & $0.5678$ & A & G & A & $4.36$  \\
	$80005^{n}$ & 04:07:48.6 & -12:11:52 & $15.5$ & $70$ & $21.54 \pm 0.00$ & $0.5656$ & A & G & A & $1.01$  \\
	$1883^{}$ & 04:07:48.3 & -12:11:21 & $15.8$ & $78$ & $21.69 \pm 0.12$ & $0.6883$     & A & G & E & $1.69$  \\
	$90002^{}$ & 04:07:49.5 & -12:11:41 & $16.3$ & $74$ & $22.10 \pm 9.99$ & $0.5715$   & A & G & A & $0.63$  \\
	$1920^{n}$ & 04:07:47.4 & -12:11:26 & $18.5$ & $96$ & $22.77 \pm 0.19$ & $0.7797$   & A & G & A & $0.97$  \\
	$1862^{}$ & 04:07:49.1 & -12:11:21 & $18.5$ & $78$ & $22.63 \pm 0.17$ & $0.4942$     & A & G & E & $0.25$  \\
	$1866^{}$ & 04:07:48.8 & -12:11:58 & $22.0$ & $100$ & $21.40 \pm 0.10$ & $0.5713$   & A & G & E & $1.19$  \\
	$80010^{n}$ & 04:07:49.8 & -12:11:48 & $23.1$ & $-1$ & $21.70 \pm 0.00$ & $1.4657$  & B & G & E & $-1.00$  \\
	$1820^{n}$ & 04:07:49.9 & -12:11:49 & $24.8$ & $113$ & $21.75 \pm 0.13$ & $0.5686$  & A & G & A & $0.85$  \\
	$1854^{}$ & 04:07:49.2 & -12:12:04 & $29.6$ & $136$ & $21.23 \pm 0.09$ & $0.5779$   & A & G & E & $1.44$  \\
	\hline
	\multicolumn{11}{l}{\bf Notes} \\
	\multicolumn{11}{l}{$^\mathrm{a}$ ID, coordinates, and $R$-band photometry from \cite{Prochaska:2006} except for objects with IDs 800\#\# or 900\#\# } \\
	\multicolumn{11}{l}{\ \ \, which were added from \cite{Ellingson:1994} or based on visual inspection of the DuPont/WFCCD or {\it HST} images.} \\
	\multicolumn{11}{l}{$^\mathrm{b}$ Given a value of $-1$ when not available due to lack of a secure redshift or when not applicable.} \\
	\multicolumn{11}{l}{$^\mathrm{c}$ Redshift and classification quality: A$\rightarrow$ secure ($\geq2$ features), B$\rightarrow$ 1 feature, C$ \rightarrow$ observed but no features, $N\rightarrow$ not observed.} \\
	\multicolumn{11}{l}{$^\mathrm{d}$ Object classification: Q$\rightarrow$ quasar, G$\rightarrow$ galaxy, S$\rightarrow$ star, U$\rightarrow$ unknown.} \\
	\multicolumn{11}{l}{$^\mathrm{e}$ Galaxy classification: E$\rightarrow$ emission-line dominated, A$\rightarrow$ absorption-line dominated, $N\rightarrow$ n/a.} \\
	\multicolumn{11}{l}{$^\mathrm{f}$ Measured from $R$-band photometry as discussed in text.} \\
	\multicolumn{11}{l}{$^\mathrm{n}$ Redshift and classification from new data presented in this paper.} \\
	\end{tabular}
\end{table*}

The new IMACS/LDSS3 observations include 225 new galaxies without
previously known redshifts, providing an unprecedented level of
completeness in the field when combined with previous surveys.
Specifically, the survey is 100\% ($\approx90$\%) complete for
galaxies brighter than $R=22$ ($R=23$) within $1'$ of the quasar
sightline. The relevant figure-of-merit, however, is the completeness
as a function of galaxy luminosity, physical impact parameter, and
redshift which is shown in Figure \ref{figure:completeness} along with
a galaxy-redshift histogram\footnote{ We estimated the completeness
  function using the observed completeness as a function of magnitude
  and angular separation from the quasar sightline and smoothed with a
  $\Delta\,z=0.1$ boxcar function to remove fluctuations due to small
  number statistics.}.  The survey is 100\% complete for $L>0.1\,L_*$
galaxies at impact parameters less than $\rho=100$ kpc and $>90\%$
complete at $\rho<200$ kpc at all redshifts $z<z_{\rm QSO}$. The
completeness level decreases somewhat at larger impact parameters due
to our targeting priority, but even at $\rho<500$ kpc, the survey is
$\gtrsim75$\% complete for $L>0.1\,L_*$ galaxies at $z<z_{\rm
  QSO}$. The high level of completeness is visually captured by the
{\it HST} image of the field labelled with galaxy redshifts (Figure
\ref{figure:hst}).

We note that in addition to the many foreground galaxies
spectroscopically identified near the quasar line-of-sight, there exists
a clear overdensity of galaxies at the redshift of the quasar.  The
galaxy catalog is therefore also valuable for studying AGN fueling
(e.g.\ Ellingson \& Yee 1994).

\begin{figure*}
	\centering
	\includegraphics[angle=0,scale=0.65]{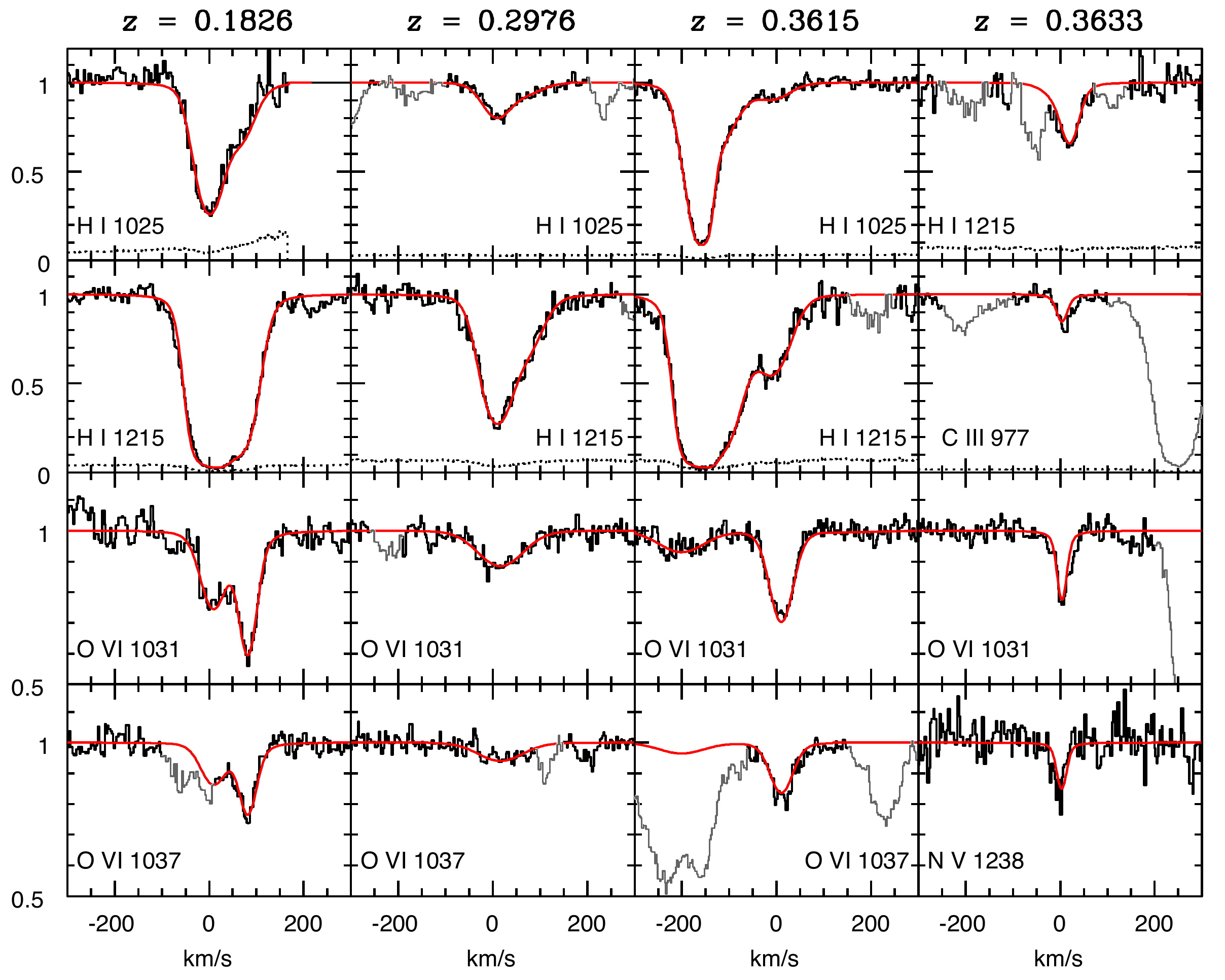}
	\caption{Absorption profiles of two O\,VI absorbers and a
          tentative detection (left three columns) and the N\,V
          doublet detected in a known metal-line absorber (right column) in
          the high-quality COS spectrum, along with the associated
          H\,I \lya\ and/or Ly$\beta$ transition. The continuum normalized spectrum is shown
          in black solid line and the associated 1-$\sigma$ errors are
          plotted in black dotted line.  The best-fit model profiles
          from a Voigt profile analysis (\S\,\ref{section:new_abs}) are shown in
          red. Spectral regions contaminated by other absorption
          systems are plotted in gray. We do not plot data at
          velocities greater than $\Delta v > +120$ \kms\ for the
          Ly$\beta$ line of the $z=0.1826$ absorber because in this
          range, the quasar flux is nearly completely attenuated by Milky
          Way Ly$\alpha$ absorption.  We consider the O\,VI absorber
          at $z=0.2977$ a tentative detection, because the observed depth of
          the $\lambda 1037$ member appears shallower than the expectation of
          the doublet based on the fit of the $\lambda 1031$ member. 
          Note that the bottom two rows are shown with a partial $y$-range from 
          $0.5$ to $1.2$ to improve the visibility of weak features. This partial $y$-range prevents the errors from being displayed in the bottom two rows.}
	\label{figure:newAbsorbers}
\end{figure*}

\section{New measurements of absorption-line systems}
\label{section:new_abs}

The higher $S/N$ of the new COS data ensured both improved
measurements of previously known transitions and new detections along
the sightline.  We conducted a systematic search of new absorption
features in the COS spectra of PKS\,0405$-$123 and identified
three features that were previously unknown.  These include: (1)
 a tentative detection of a new O\,VI absorber at $z=0.2977$, 
 (2) a new O\,VI component at $\Delta\,v=+170$
\kms\ from a previously known metal-line absorber at $z=0.3608$, and (3)
N\,V absorption associated with a known O\,VI system at $z = 0.3633$.
Combining previous results with the new findings yielded a sample of
seven O\,VI absorbers and 12 individual components found at
$z=0.0918-0.495$ along the sightline toward PKS\,0405$-$123.

\begin{table}
  \caption{Summary of line properties of known O\,VI absorbers}
	\label{table:oviabsorbers}
	\begin{tabular}{lcccc}
	\hline
	\hline
	 & $\Delta\,v$ & $b$&  & \\
	Element & (\kms) & $($\kms$)$& $\log N\,/\,{\rm cm^{-2}}$ & References$^{\rm a}$ \\
	\hline
	\multicolumn{5}{c}{$z_{\rm sys} = 0.09180$}\\
	\hline
	H\,I & $0$ & $38\pm2$ & $14.52 \pm 0.05$ & 1 \\
	C\,IV  & $0$ & $$ & $<12.90$ & 2\\
	N\,V  & $0$ & $$ & $<12.67$ & 2\\
	O\,VI & $+20$ & $$ & $13.83 \pm 0.04$ & 1\\
	Si\,IV  & $0$ & $$ & $<12.66$ & 2\\
	\hline
	\multicolumn{5}{c}{$z_{\rm sys} = 0.09658$}\\
	\hline
	H\,I  & $0$ & $40 \pm 2$ & $14.65 \pm 0.05$ & 1 \\
	C\,IV & $0$ & $$ & $<12.90$ & 2\\
	N\,V  & $0$ & $$ & $<12.97$ & 2\\
	O\,VI  & $0$ & $$ & $13.70 \pm 0.20$ & 1 \\
	Si\,IV  & $0$ & $$ & $<12.64$ & 2\\
	\hline
	\multicolumn{5}{c}{$z_{\rm sys} = 0.16711$}\\
	\hline
	 H\,I  & $-11 \pm 3$ & $$ & $15.50$ & 1, 3 \\
	 H\,I & $0 \pm 5$ & $$ & $16.35$ & 1, 3 \\
	 O\,VI & $-266 \pm 5$ & $52 \pm 2$ & $13.90 \pm 0.03$ & 3 \\
	 O\,VI & $-35 \pm 2$ & $$ & $14.71 \pm 0.01$ & 3 \\
	\hline
	\multicolumn{5}{c}{$z_{\rm sys} = 0.18259$}\\
	\hline
	H\,I & $0 \pm 4$ & $33 \pm 1$ & $14.69 \pm 0.01$ &  2\\ 
	H\,I & $+68 \pm 4$ & $36 \pm 2$ & $14.09 \pm 0.03$ &  2\\
	N\,V  & $0$ & $$ & $<12.83$ & 2\\
	O\,VI  & $+9 \pm 4$ & $32 \pm 3$ & $13.75 \pm 0.02$ & 2\\
	O\,VI  & $+82 \pm 4$ & $21 \pm 1$ & $13.88 \pm 0.02$ & 2\\
	Si\,IV  & $0$ & $$ & $<12.77$ & 2\\
	\hline
	\multicolumn{5}{c}{$z_{\rm sys} = 0.29762$}\\
	\hline
	H\,I & $0 \pm 5$ & $38 \pm 3$ & $13.89 \pm 0.05$ &  2\\
	H\,I & $+65 \pm 15$ & $42 \pm 12$ & $13.34 \pm 0.19$ &  2\\
	C\,III  & $0$ & $$ & $<12.05$ & 2\\
	N\,V  & $0$ & $$ & $<13.13$ & 2\\
	O\,VI$^{\rm b}$ & $+8 \pm 4$ & $63 \pm 4$ & $13.61 \pm 0.02$ &  2\\
	\hline
	\multicolumn{5}{c}{$z_{\rm sys} = 0.36078$}\\
	\hline
	H\,I & $-24 \pm 14$ & $26 \pm 8$ & $14.3 \pm 0.3$ &  2\\ 
	H\,I & $0 \pm 3$ & $16 \pm 1$ & $15.06 \pm 0.05$ &  2\\
	H\,I & $+27 \pm 8$ & $54 \pm 6$ & $14.29 \pm 0.08$ &  2\\
	H\,I & $+151 \pm 4$ & $45 \pm 4$ & $13.56 \pm 0.04$ &  2\\
	H\,I$^{\rm c}$ & $+557 \pm 3$ & $41 \pm 10$ & $12.75 \pm 0.41$ &  2\\
	H\,I & $+575 \pm 3$ & $22 \pm 5$ & $13.11 \pm 0.14$ &  2\\
	C\,III$$  & $+168$ & $$ & $<12.5$ & 2\\
	C\,III$^{\rm c}$  & $+557 \pm 3$ & $12$ & $12.48\pm 0.07$ & 2\\
	N\,V$$  & $+168$ & $$ & $<13.1$ & 2\\
	N\,V$^{\rm c}$  & $+557 \pm 3$ & $11$ & $13.01 \pm 0.08$ & 2\\
	O\,VI$^{\rm b}$ & $-44 \pm 7$ & $64 \pm 9$ & $13.38 \pm 0.05$ &  2\\
	O\,VI & $+168 \pm 3$ & $30 \pm 1$ & $13.80 \pm 0.01$ &  2\\
	O\,VI$^{\rm c}$ & $+557 \pm 3$ & $10$ & $13.36 \pm 0.05$ &  2\\
	\hline
	\multicolumn{5}{c}{$z_{\rm sys} = 0.49510$}\\
	\hline
	H\,I & $0$ & $53 \pm 1$ & $14.09 \pm 0.03$ &  4 \\
	H\,I & $+93$ & $21 \pm 2$ & $13.44 \pm 0.09$ &  4 \\
	O\,VI & $-3$ & $35 \pm 1$ & $14.29 \pm 0.02$ & 4\\
	O\,VI & $+48$ & $19 \pm 2$ & $13.80 \pm 0.07$ & 4\\
	\hline
        \multicolumn{5}{l}{$^{\rm a}$\,1: \protect{\citealt{Prochaska:2004}}; 2: this work;} \\
        \multicolumn{5}{l}{\ \,\,3: \protect{\citealt{Savage:2010}}; 4: \protect{\citealt{Narayanan:2011}}.}\\
        \multicolumn{5}{l}{$^{\rm b}$\,Tentative detection due to contamination of the $\lambda\,1037$ member.} \\
         \multicolumn{5}{l}{$^{\rm c}$\,Measurements obtained with a simultaneous fit to the transitions} \\
        \multicolumn{5}{l}{$^{\rm }$\ \,\,assuming that these absorbers originate in the same gas.} \\
        \end{tabular}
\end{table}

We measured the line profiles of both new and known O\,VI absorbers,
as well as their associated H\,I and other metal transitions based on
a Voigt profile analysis using the VPFIT
package\footnote{http://www.ast.cam.ac.uk/$\sim$rfc/vpfit.html}
\citep{Carswell:1987} and the empirical COS line spread function
(LSF)\footnote{http://www.stsci.edu/hst/cos/performance/spectral\_resolution/}.  The COS LSF exhibits broad wings which
contain up to 40\% of the flux \citep{Ghavamian:2009}.
The use of the empirical LSF in the
Voigt profile fitting is therefore necessary in order to properly
account for the absorption that falls in the wings of the LSF.  In
cases of line blending, we employed a simultaneous fit of a minimum
number of separate components that are necessary to produce a
reasonable $\chi^2$ value.  For non-detections, we calculated the
3-$\sigma$ rest-frame equivalent width limit over a $300$ \kms\
spectral window (significantly broader than the wings of the COS LSF)
and converted this to the corresponding 3-$\sigma$ limit in column
density assuming that the gas is optically thin.  The results are
summarized in Table \ref{table:oviabsorbers}, where we list for each
species the velocity offset $\Delta\,v$ relative to the systemic
redshift of each absorber $z_{\rm sys}$ as defined by the dominant
H\,I component, the Doppler parameter $b$, and the best-fit column
density.  We also include in Table \ref{table:oviabsorbers}
measurements from the literature for completeness.  Here we briefly
describe the properties of newly detected absorption features.

At $z=0.1826$ and $0.1829$, two O\,VI components were found
in previous searches \citep{Prochaska:2004, Thom:2008a}.
We confirm both detections and remeasure the absorber properties
with a simultaneous fit of both components in the COS data.
The lower-redshift O\,VI component is characterized by an
O\,VI column density $\log\,N({\rm O\,VI})= 13.75 \pm 0.02$ and
Doppler width $b = 32 \pm 3$ \kms, and is merely $\approx 9$ \kms\
offset from a previously identified H\,I absorption component with
$\log\,N({\rm H\,I})=14.69 \pm 0.01$ and $b = 33\pm 1$ \kms.  
The higher-redshift O\,VI component is characterized by $\log\,N({\rm O\,VI}) = 13.88 \pm
0.02$ and $b=21 \pm 1$ \kms.  No
other ions have been detected with a 3-$\sigma$ upper limit of
$\log\,N<12.8$ for both N\,V and Si\,IV absorption.  \cite{Prochaska:2004} 
reported an upper limit of $\log\,N<12.4$ for C\,III
absorption.  The best-fit Voigt profiles of the O\,VI doublet along
with those of \lya\ and \lyb\ absorption are presented in the left
column of Figure \ref{figure:newAbsorbers}).

\begin{table}
  \caption{Summary of line properties of strong \lya\ absorbers with no detectable O\,VI absorption}
	\label{table:hiabsorbers}
	\begin{tabular}{lcccc}
	\hline
	\hline
	 & $\Delta v$ & $b$& & \\
	Element & (\kms) & $($\kms$)$& $\log\,N\,/\,{\rm cm^{-2}}$ & References$^{\rm a}$ \\
	\hline
	\multicolumn{5}{c}{$z_{\rm sys} = 0.08139$ }\\
	\hline
	H\,I & $0$ & $54 \pm 4$ & $13.79 \pm 0.02$ &  1\\
	O\,VI & $0$ & $$ & $<13.62$ & 2\\
	\hline
	\multicolumn{5}{c}{$z_{\rm sys} = 0.13233$}\\ 
	\hline
	H\,I & $0$ & $22 \pm 2$ & $13.64 \pm 0.03$ &  1\\
	H\,I & $+193$ & $32 \pm 6$ & $13.29 \pm 0.06$ &  1\\
	C\,IV & $0$ & $$ & $<12.94$ & 3\\
	N\,V & $0$ & $$ & $<12.97$ & 3\\
	O\,VI & $0$ & $$ & $<12.95$ & 3\\
	Si\,IV & $0$ & $$ & $<12.83$ & 3\\
	\hline
	\multicolumn{5}{c}{$z_{\rm sys} = 0.15304$} \\
	\hline
	 H\,I & $-208$ & $22 \pm 2$ & $13.54 \pm 0.04$ &  1\\
	 H\,I & $0$ & $46 \pm 3$ & $13.80 \pm 0.03$ &  1\\
	 C\,IV & $0$ & $$ & $<13.30$ & 3\\
	 N\,V & $0$ & $$ & $<12.71$ & 3\\
	 O\,VI & $0$ & $$ & $<13.14$ & 3\\
	 Si\,IV & $0$ & $$ & $<12.56$ & 3\\
	\hline
	\multicolumn{5}{c}{$z_{\rm sys} = 0.16121$} \\ 
	\hline
	 H\,I & $0$ & $54 \pm 8$ & $13.71 \pm 0.04$ &  1\\
	 H\,I & $+75$ & $18 \pm 4$ & $13.27 \pm 0.09$ &  1\\
	 N\,V & $0$ & $$ & $<13.11$ & 3\\
	 O\,VI & $0$ & $$ & $<12.85$ & 3\\
	 Si\,IV & $0$ & $$ & $<12.42$ & 3\\
	\hline
	\multicolumn{5}{c}{$z_{\rm sys} = 0.17876$}\\ 
	\hline
	 H\,I & $0$ & $55 \pm 7$ & $13.61 \pm 0.04$ &  1\\
	 C\,III & $0$ & $$ & $<12.37$ & 3\\
	 N\,V & $0$ & $$ & $<13.13$ & 3\\
	 O\,VI & $0$ & $$ & $<13.16$ & 3\\
	\hline
	\multicolumn{5}{c}{$z_{\rm sys} = 0.24554$} \\
	\hline
	 H\,I & $-99$ & $54 \pm 24$ & $13.23 \pm 0.11$ & 1 \\
	 H\,I & $0$ & $23 \pm 2$   & $13.69 \pm 0.03$ & 1 \\
	 O\,VI & $0$ & $$ & $<12.89$ & 3\\
	 Si\,IV & $0$ & $$ & $<12.53$ & 3\\
	\hline
	\multicolumn{5}{c}{$z_{\rm sys} = 0.33402$} \\
	\hline
	 H\,I & $0$ & $30 \pm 2$ & $13.82 \pm 0.03$ & 1 \\
	 N\,V & $0$ & $$ & $<13.04$ & 3\\
	 O\,VI & $0$ & $$ & $<13.12$ & 3\\
	 \hline
	\multicolumn{5}{c}{$z_{\rm sys} = 0.35099$} \\ 
	\hline
	 H\,I & $0$ & $38 \pm 2$ & $14.25 \pm 0.03$ & 1 \\
	 H\,I & $+115$ & $25 \pm 5$ & $13.53 \pm 0.05$ & 1 \\
	 H\,I & $+251$ & $31 \pm 5$ & $13.57 \pm 0.05$ & 1 \\
	 C\,III & $0$ & $$ & $<12.14$ & 3\\
	 N\,V & $0$ & $$ & $<13.05$ & 3\\
	 O\,VI & $0$ & $$ & $<13.00$ & 3\\
	\hline
	\multicolumn{5}{c}{$z_{\rm sys} = 0.40571$} \\
	\hline
	 H\,I & $0$ & $33 \pm 2$ & $14.98 \pm 0.02$ & 1 \\
	 C\,III & $0$ & $$ & $<12.00$ & 3\\
	 N\,V & $0$ & $$ & $<13.10$ & 3\\
	 O\,VI & $0$ & $$ & $<13.28$ & 3\\
	\hline
	\multicolumn{5}{l}{$^{\rm a}$\,1: \protect{\citealt{Lehner:2007}}; 2: from FUSE data published in }\\
	\multicolumn{5}{l}{\ \,\,\protect{\citealt{Prochaska:2004}}; 3: this work; 4: \protect{\citealt{Williger:2006}}}\\
	\end{tabular}
\end{table}
\addtocounter{table}{-1}
\begin{table}
	\caption{Continued.}
	\begin{tabular}{lcccc}
	\hline
	\hline
	 & $\Delta\,v $ & $b$&  & \\
	Element & (\kms) & $($\kms$)$& $\log\,N\,/\,{\rm cm^{-2}}$ & References$^{\rm a}$ \\
	\hline
	\multicolumn{5}{c}{$z_{\rm sys} = 0.40886$}\\
	\hline
	 H\,I & $0$ & $40 \pm 2$ & $14.38 \pm 0.03$ & 1 \\
	 H\,I & $+149$ & $26 \pm 6$ & $13.58 \pm 0.07$ & 1 \\
	 N\,V & $0$ & $$ & $<13.12$ & 3\\
	 O\,VI & $0$ & $$ & $<13.28$ & 3\\
	\hline
	\multicolumn{5}{c}{$z_{\rm sys} = 0.53830$} \\
	\hline
	 H\,I & $0$ & $23 \pm 4$ & $14.22 \pm 0.06 $ & 4  \\
	 C\,III & $0$ & $$ & $<12.35$ & 3\\
	 O\,VI & $0$ & $$ & $<13.26$ & 3\\
	\hline
	\multicolumn{5}{l}{$^{\rm a}$\,1: \protect{\citealt{Lehner:2007}}; 2: from FUSE data published in }\\
	\multicolumn{5}{l}{\ \,\,\protect{\citealt{Prochaska:2004}}; 3: this work; 4: \protect{\citealt{Williger:2006}}}\\
	\end{tabular}
\end{table}

At $z=0.2977$, we report a tentative detection of an O\,VI absorber at
$\Delta\,v\approx 8$ \kms\ from a previously identified \lya\ absorber.  The
O\,VI absorption profile is relatively broad with $\log\,N({\rm
  O\,VI}) = 13.61 \pm 0.02$ and $b\approx 60$ \kms.  The
$\lambda\,1037$ member is detected at a $\approx 7$-$\sigma$ level of
significance, but the model slightly over-predicts the absorption
strength of this weaker member which appears to be contaminated 
by other absorption features.  No other metal absorption is detected
in the COS spectrum.  Our Voigt profile analysis shows that the \lya\
transition is best described by two components separated by
$\Delta\,v=65$ \kms, with one component containing $\log\,N({\rm
  H\,I})=13.89\pm 0.05$ at $z=0.2976$ and the other containing
$\log\,N({\rm H\,I})=13.34\pm 0.2$ at $z=0.2979$.  The best-fit Voigt
profiles of the O\,VI doublet along with those of \lya\ and \lyb\
absorption are presented in the middle-left column of Figure
\ref{figure:newAbsorbers}).

\begin{figure*}
	\centering
	\includegraphics[scale=0.6, angle=0, trim=0in 0in 0in 0.05in]{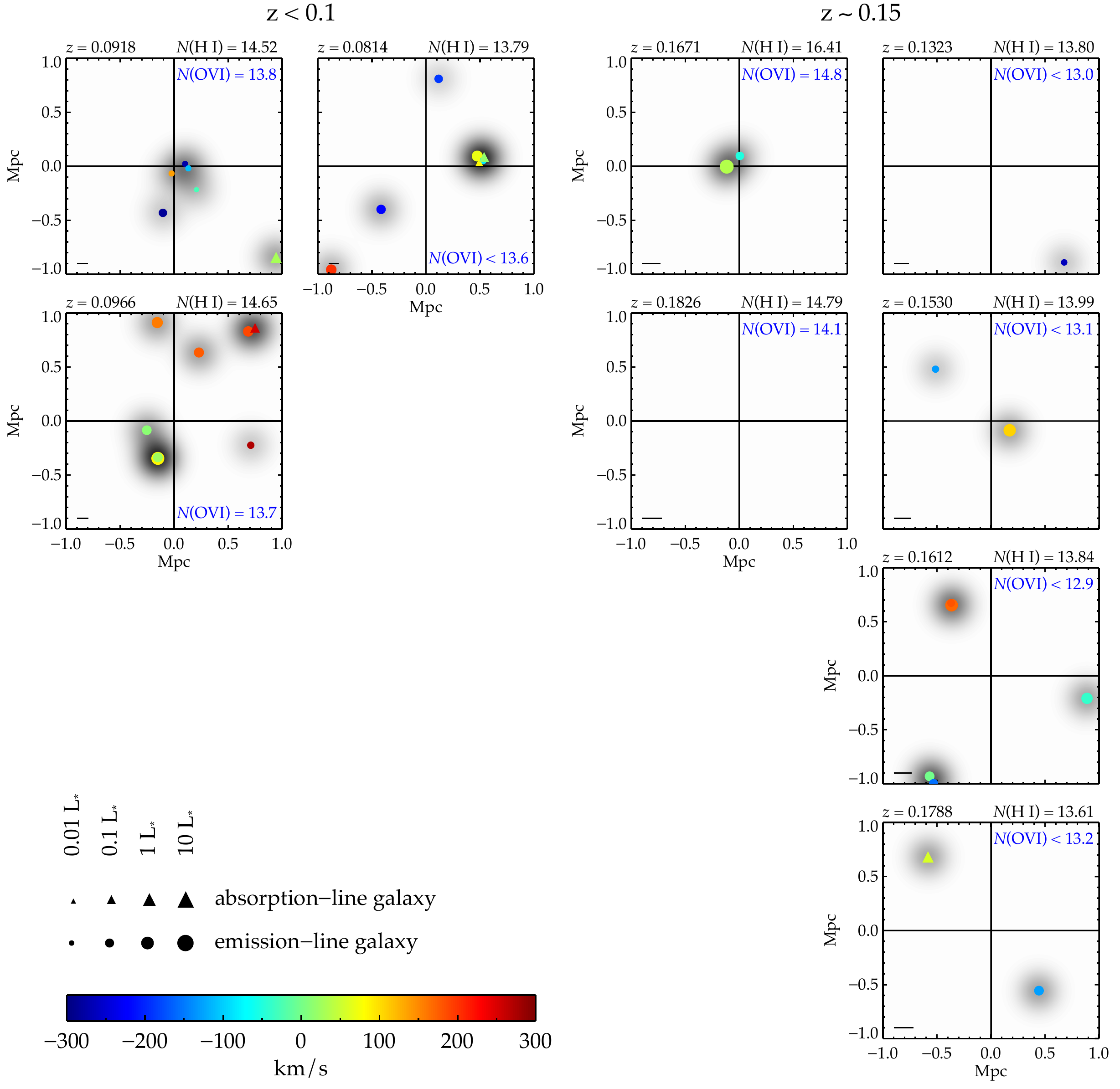}
	\caption{The galaxy environments of O\,VI absorbers and strong
          Ly$\alpha$ absorbers of $\log\,N({\rm H\,I})>13.6$ with no
          detectable O\,VI (designated as ``\lya-only'' systems).
          Considering the increasing survey incompleteness of
          faint ($<0.1\,L_*$) galaxies with increasing redshift, we
          separate the galaxy--absorber sample into four redshift
          bins.  For each redshift bin, we show the spatial and
          velocity distributions of galaxies around O\,VI absorbers in
          the left column and \lya-only absorbers in the right column.
          Each panel is centered at the quasar, while positions
          of galaxies with projected line-of-sight velocity $|\Delta v|<300 $ \kms\ of the absorber
           are marked with circles for emission-line galaxies and
          triangles for absorption-line galaxies. The symbols are
          color-coded to indicate the line-of-sight velocity between
          each galaxy and the absorber.  
          The symbol size specifies
          galaxy luminosity as shown in the figure legend.  To help
          visualize the surface density of surrounding galaxies, we
          also introduce a gray-scale showing a luminosity-weighted
          galaxy surface density where each galaxy is represented by a
          Gaussian with ${\rm FWHM}=300$ kpc.  The luminosity weighting
          assigns $L\geq L_*$ galaxies a peak of 1, $0.1\leq L<L_*$
          galaxies a peak of $0.7$ and $L<0.1\,L_*$ galaxies a peak of
          $0.4$.}
	\label{figure:environment}
\end{figure*}

\setcounter{figure}{4}
\begin{figure*}
	\includegraphics[scale=0.49, angle=0]{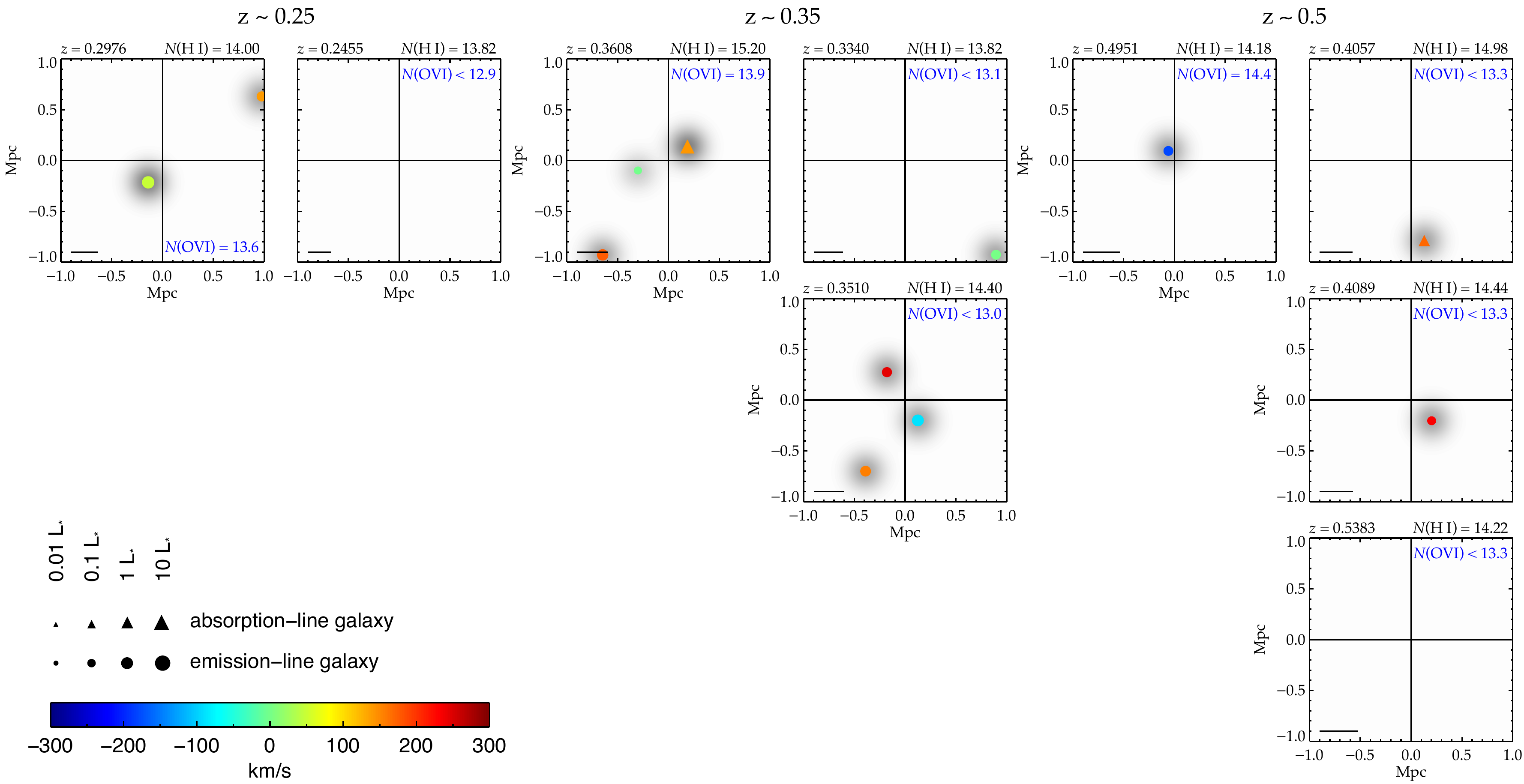}
	\caption{continued}
	\label{figure:environments}
\end{figure*}

At $z = 0.3615$, we also report the detection of a new O\,VI component
with $\log\,N({\rm O\,VI}) = 13.80 \pm 0.01$ and $b=30 \pm 1$ \kms\ at
$\Delta\,v = +170$ \kms\ from a previously known absorption complex at
$z = 0.3608$ (e.g.\ Prochaska \etal\ 2004).  Similar to the 
O\,VI absorber at $z=0.1829$, we do not detect
additional metal-line systems associated with this new O\,VI
component.  There is possible O\,VI absorption at $z\approx0.3608$
with $\log\,N({\rm O\,VI}) =13.38\pm0.05$ detected in the COS
spectrum, but it cannot be confirmed due to contaminating features at
the location of the second doublet member.  The best-fit Voigt
profiles of the O\,VI doublet along with those of \lya\ and \lyb\
absorption are presented in the middle-right column of Figure
\ref{figure:newAbsorbers}).

Finally, the O\,VI absorption system previously identified at
$z=0.3633$ with $\log\,N({\rm O\,VI}) =13.36\pm0.05$ and $b=10 \pm 1$
\kms\ displays possible presence of associated N\,V doublet in the new
COS spectrum.  Prochaska \etal\ (2004) identified H\,I, O\,IV, N\,IV,
and C\,III absorption associated with this O\,VI absorber, although
the H\,I absorbing component is $\Delta\,v=+20$ \kms\ away from the
O\,VI absorber.  We confirm the presence of C\,III absorption with
$\log\,N({\rm C\,III}) =12.48\pm0.07$ and detect N\,V with $N({\rm
  N\,V}) = 13.01 \pm 0.08$ at the redshift of the O\,VI absorber.  The
measurements were obtained based on a simultaneous fit of Voigt
profiles to all three transitions, assuming that these ions originate in
the same gas.  Allowing independent fits to the line centroids of N\,V
and C\,III does not improve the model-fit.
The best-fit Voigt profiles of \lya\,, C\,III $\lambda\,977$, O\,VI $\lambda\,1031$ 
and N\,V $\lambda\,1238$ are presented in the right column of Figure
\ref{figure:newAbsorbers}).  Both O$^{5+}$ and N$^{4+}$ are highly
ionized species, whereas C$^{2+}$ is at a much lower ionization state.
The presence of these ions together strongly support a
photo-ionization scenario for the absorbing gas (see \citealt{Prochaska:2004}).
Based on the observed relative abundance of C$^{2+}$ and
O$^{5+}$, \citealt{Prochaska:2004} derived an ionization parameter of
$\log\,U=-1.4\pm 0.2$ and a minimum metallicity of solar for the absorbing
 gas assuming that all of the observed $N({\rm H\,I})$ is associated with 
 the O\,VI absorbing gas. 
 
Including the tentative detection of an O\,VI absorber at $z=0.2977$,
we have a total of seven O\,VI absorbers along the sightline toward
PKS\,0405$-$123.  In addition, this sightline contains additional 11
strong \lya\ absorption systems of $\log\,N({\rm H\,I})>13.6$
\citep{Williger:2006,Lehner:2007}\footnote{ We associate the H\,I
  components identified by \cite{Lehner:2007} with one another
  provided $|\Delta v| < 300$ \kms.  \cite{Lehner:2007} restricted
  their study to absorbers with $z<0.5$ so we include systems
  identified by \cite{Williger:2006} at $z>0.5$ as well.  For each
  system, we adopt the total H\,I column density and the redshift of
  the strongest H\,I component.}  with no detectable absorption from
heavy ions.  The high $S/N$ of the COS spectrum enables stronger
limits \citep[cf. ][]{Prochaska:2004} on the associated O\,VI column
densities.  We define a subsample of ``\lya-only'' absorbers of these
strong \lya\ absorbers with no associated O\,VI absorption to a
sensitive upper limit.  The properties of these \lya-only systems are
summarized in Table \ref{table:hiabsorbers}.

\section{The galaxy environments of O\,VI and \lya-only Absorber}
\label{section:environments}

The highly complete survey data of faint galaxies in the field 
around PKS 0405$-$123 offer a new opportunity to re-examine 
the galaxy environments of O\,VI absorbers.  
In particular, we compare our findings with previous
results from a shallower survey which concluded that O\,VI
absorbers trace a diverse set of environments including: the halos of
individual galaxies, galaxy groups, filamentary-like structures, and
galaxy voids \citep[e.g.][]{Prochaska:2006}.  In addition, we compare
the galaxy environments of the O\,VI absorbers with those of \lya-only
absorbers which constitutes a control sample in seeking the
discriminating galactic features that result in the observed O\,VI
absorbing gas at large galactic radii. Throughout, we associate
galaxies with absorbers provided the projected line-of-sight velocity
between the absorber and the galaxy is $|\Delta v| < 300$ \kms.  All
galaxies spectroscopically identified at impact parameter $\rho<1$ Mpc and
velocity offset $|\Delta v| < 300$ \kms \, of the O\,VI and \lya-only
systems are presented in Tables \ref{table:galaxies_OVI} and
\ref{table:galaxies_HI}, respectively.

 The galactic environments of O\,VI absorbers uncovered in our survey
 are presented in Figure \ref{figure:environments}.  We find that
 O\,VI primarily traces over-dense galaxy environments with at least
 one emission-line galaxy found within $\rho\approx 300$ kpc.
 Specifically, the O\,VI absorber at $z=0.0918$ was originally
 attributed to filamentary structure connecting three galaxy groups at
 $\rho=1-3$ Mpc (Prochaska \etal\ 2006).  However, we have uncovered
 four dwarf, emission-line galaxies at $\rho \leq 300$ kpc, the
 closest of which is at $\rho=70$ kpc and $\Delta v = +140$ \kms.
 Similarly, the absorbers at $z\approx 0.3608$ had been
 attributed to the intragroup medium of a group of passive galaxies
 found at $\rho=1-3$ Mpc.  Our survey has revealed a dwarf
 emission-line galaxy ($L=0.08\,L_*$ at $\rho=320$ kpc and 
 $\Delta v \approx 0$ \kms) in addition to the known massive absorption-line
 galaxy previously found at $\rho=230$ kpc.  The only intriguing exception is
 the O\,VI absorption system at $z\approx0.183$, which exhibits two
 strong components separated by 70 \kms.  This O\,VI absorber does
 not have an associated galaxy at $\rho<300$ kpc.  Our galaxy survey rules
 out the presence of any galaxies of $L>0.04\,L_*$ at
 $\rho<250$ kpc and the presence of any galaxies of $L>0.3\,L_*$ at
 $\rho<1$ Mpc.  The lack of galaxies found in the vicinity
 suggests that this metal-enriched absorber resides in an apparent
 void.  Finally, the O\,VI absorber at $z=0.495$ with associated
 Ne\,VIII doublet (Narayanan \etal\ 2011) is found to be associated
 with an emission-line galaxy at $\rho=110$ kpc (Chen \& Mulchaey
 2009).  Our new survey has not uncovered any additional galaxies in
 the vicinity of the absorber.  We are able to rule out the presence
 of any galaxies with $L>0.1\,L_*$ at $\rho<200$ kpc.

\begin{figure*}
	 \includegraphics[angle=0,scale=0.43]{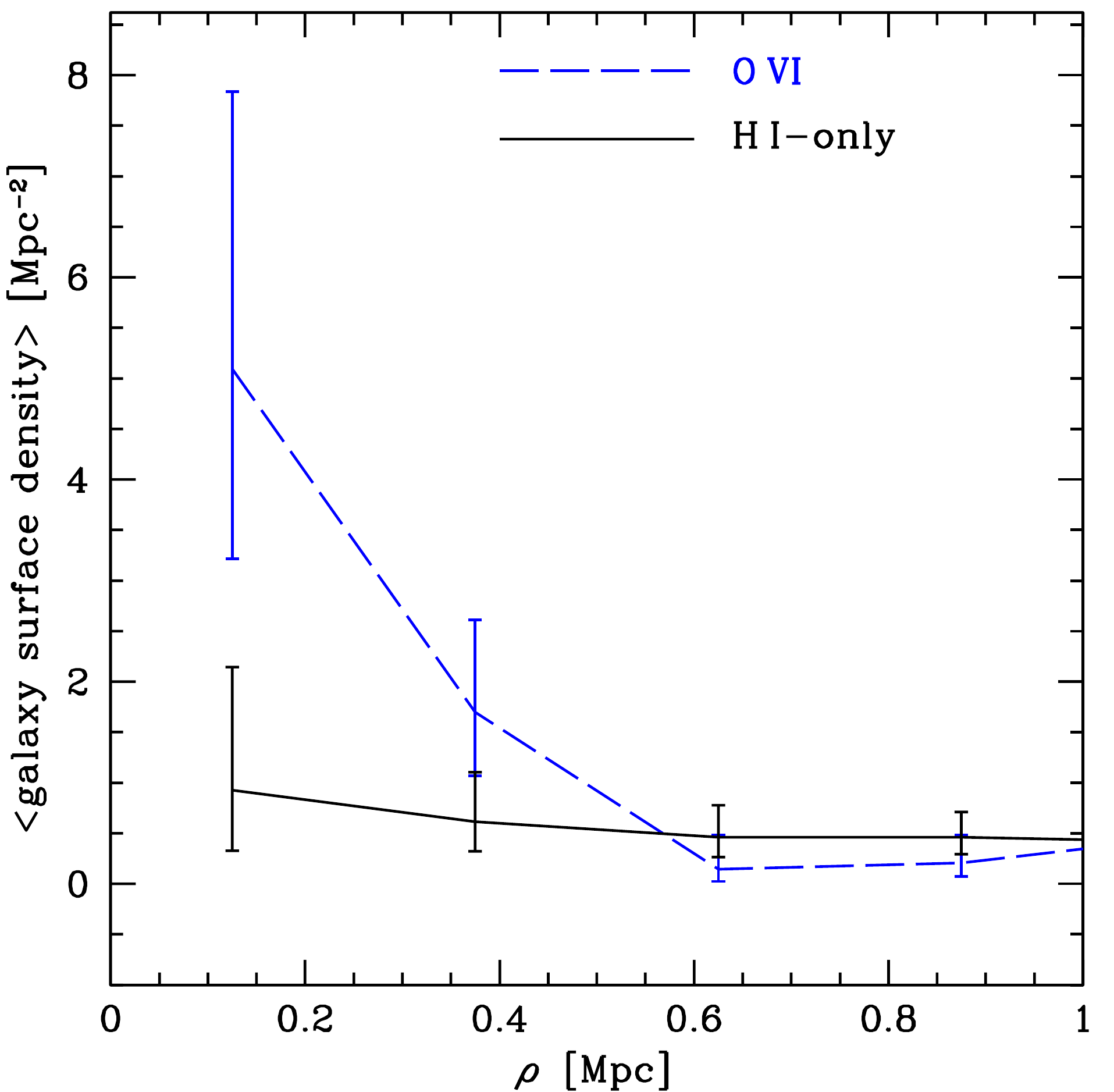} 
         \ \ \ \ \  \includegraphics[angle=0,scale=0.43, trim=0in 0.0in 0in 0in]{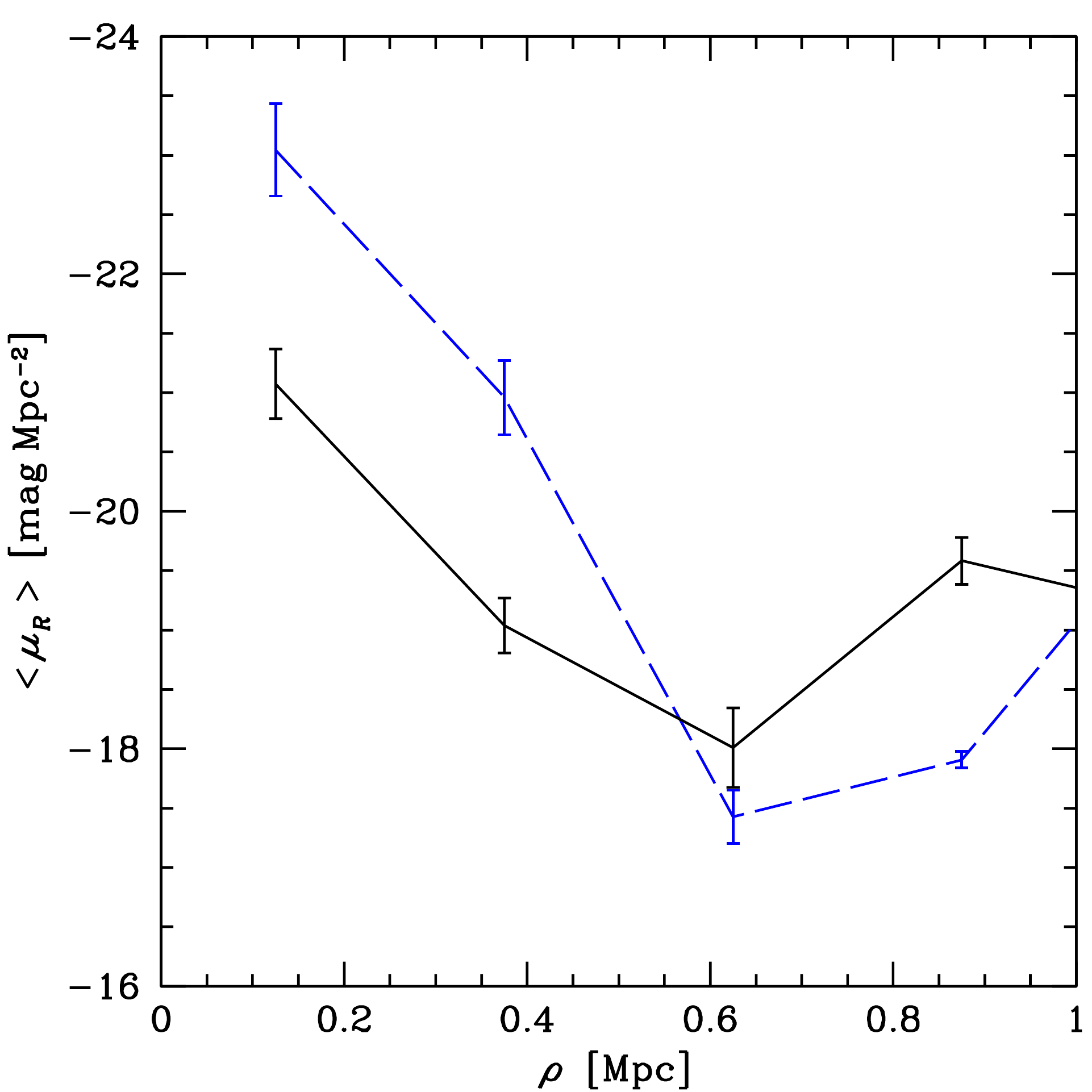}
         \caption{Mean radial profiles of the galaxy distribution around
           O\,VI and \lya-only absorbers.  The {\it left} panel shows
           the mean surface density of galaxies versus projected
           distance around O\,VI (dashed line) and \lya-only (solid
           line) absorbers, and the {\it right} panel shows the mean
           surface brightness profiles.  Galaxy density error-bars are 
           computed from Poisson confidence intervals while
           surface brightness uncertainties are computed using 
           a jackknife resampling technique to indicate uncertainties
           due to sample variance.  It is clear that there exists an
           overdensity of galaxies within $\rho\approx 500$ kpc radius of
           O\,VI absorbers, which is not seen around \lya-only
           absorbers.  Similarly, the mean galaxy surface brightness
           profile around O\,VI absorbers exhibits a steep rise by $\Delta \mu_R \approx +5\, {\rm mag \, Mpc^{-2}}$ toward
           the inner regions at $\rho\lesssim 500$ kpc, while the
           galaxy surface brightness profile around \lya-only
           absorbers remains comparatively flat with $\Delta \mu_R \approx +2\, {\rm mag \, Mpc^{-2}}$.}
	\label{figure:average}
\end{figure*}

We also present in Figure \ref{figure:environments} the galaxy
environments of O\,VI and \lya-only absorbers.  Considering the increasing
survey incompleteness of faint ($<0.1\,L_*$) galaxies with increasing
redshift, we separate the galaxy--absorber sample into four redshift
bins.  For each redshift bin, we show the spatial and velocity
distributions of galaxies around O\,VI absorbers in the left column
and \lya-only absorbers in the right column.  Each panel is centered
at the quasar, while galaxy positions are marked with circles for
emission-line galaxies and triangles for absorption-line galaxies. The
symbols are color-coded to indicate the line-of-sight velocity between
each galaxy and the absorber.  The symbol size specifies galaxy
luminosity as shown in the figure legend.  To help visualize the
surface density of surrounding galaxies, we also introduce a
gray-scale showing luminosity-weighted galaxy surface density where
each galaxy is represented by a Gaussian with ${\rm FWHM}=300$ kpc.

A qualitative finding based on Figure \ref{figure:environments} is
that a larger fraction of \lya-only absorbers appear in underdense or
relatively isolated galaxy environments in comparison to those of
O\,VI absorbers.  Considering all the absorbers together, six of the seven O\,VI
absorption systems at $z< 0.5$ along the sightline are associated with
at least one galaxy with $\rho \lesssim 300$ kpc.  Four of the six
O\,VI absorption systems at $z<0.4$ are found in a group of multiple
galaxies with $\rho \lesssim 300$ kpc. In contrast, of the 11
additional \lya-only absorbers, only three have associated galaxies
found at $\rho<300$ kpc. 

To quantify the potential difference in the observed
galactic environments between O\,VI and \lya-only absorbers, we
perform two separate tests in the following discussion.  First, we
examine the mean radial profiles of galaxy properties averaged over
all systems in each subsample.  Second, we examine the distribution of
galaxy properties within each subsample.  In both tests, the
comparisons are based on both the surface density of the galaxies and
the surface brightness of star light.

\subsection{Do O\,VI and \lya-only absorbers share similar azimuthally averaged galaxy distributions out to 1 Mpc?}
\label{section:radial}
To determine whether or not O\,VI and \lya-only absorbers
occur in similar galaxy environments, we measure the mean radial
profile of the galaxy distribution around these absorbers by first
stacking the observed 2D distribution of galaxies around individual
absorbers shown in Figure \ref{figure:environment} and then computing an azimuthal average in
annuli with increasing radius.  The left panel of Figure \ref{figure:average} displays
the mean galaxy surface density profiles around O\,VI absorbers (dashed line)
and \lya-only absorbers (solid line).  Error-bars show our estimate of
uncertainties due to counting statistics \citep[e.g.][]{Gehrels:1986}.  
It is clear that there exists an overdensity of galaxies within a $\rho\approx 500$ 
kpc radius of O\,VI absorbers, which is not seen around \lya-only absorbers.

Such distinction between O\,VI and \lya-only absorbers is also seen in
the observed mean surface brightness profiles around these absorbers
(the right panel of Figure \ref{figure:average}).  In this case, error-bars are computed
using a jackknife resampling technique to estimate uncertainties due
to sample variance.  While the mean galaxy surface brightness profile
around O\,VI absorbers exhibits a steep rise of $\Delta \mu_R \approx +5\, {\rm mag \, Mpc^{-2}}$ 
 toward the inner regions at $\rho\lesssim 500$ kpc, only a mild increase of $\Delta \mu_R \approx +2\, {\rm mag \, Mpc^{-2}}$ 
is seen in the galaxy
surface brightness profile around \lya-only absorbers.

\subsection{Do O\,VI and \lya-only absorbers share similar galaxy distribution functions in the inner 500 kpc?}

\begin{figure*}
	\includegraphics[angle=0,scale=0.43]{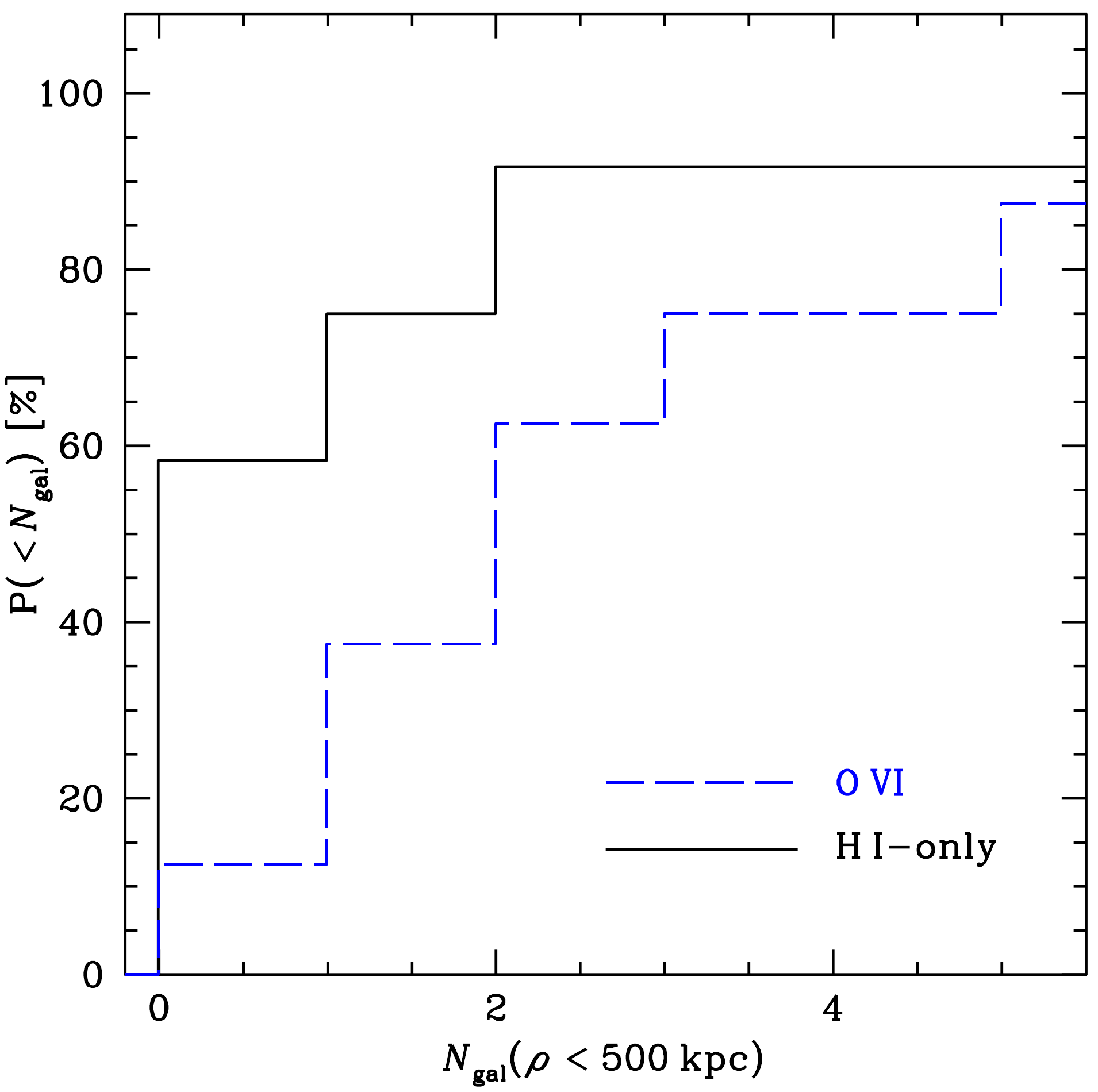} 
        \ \ \ \ \  \includegraphics[angle=0,scale=0.43]{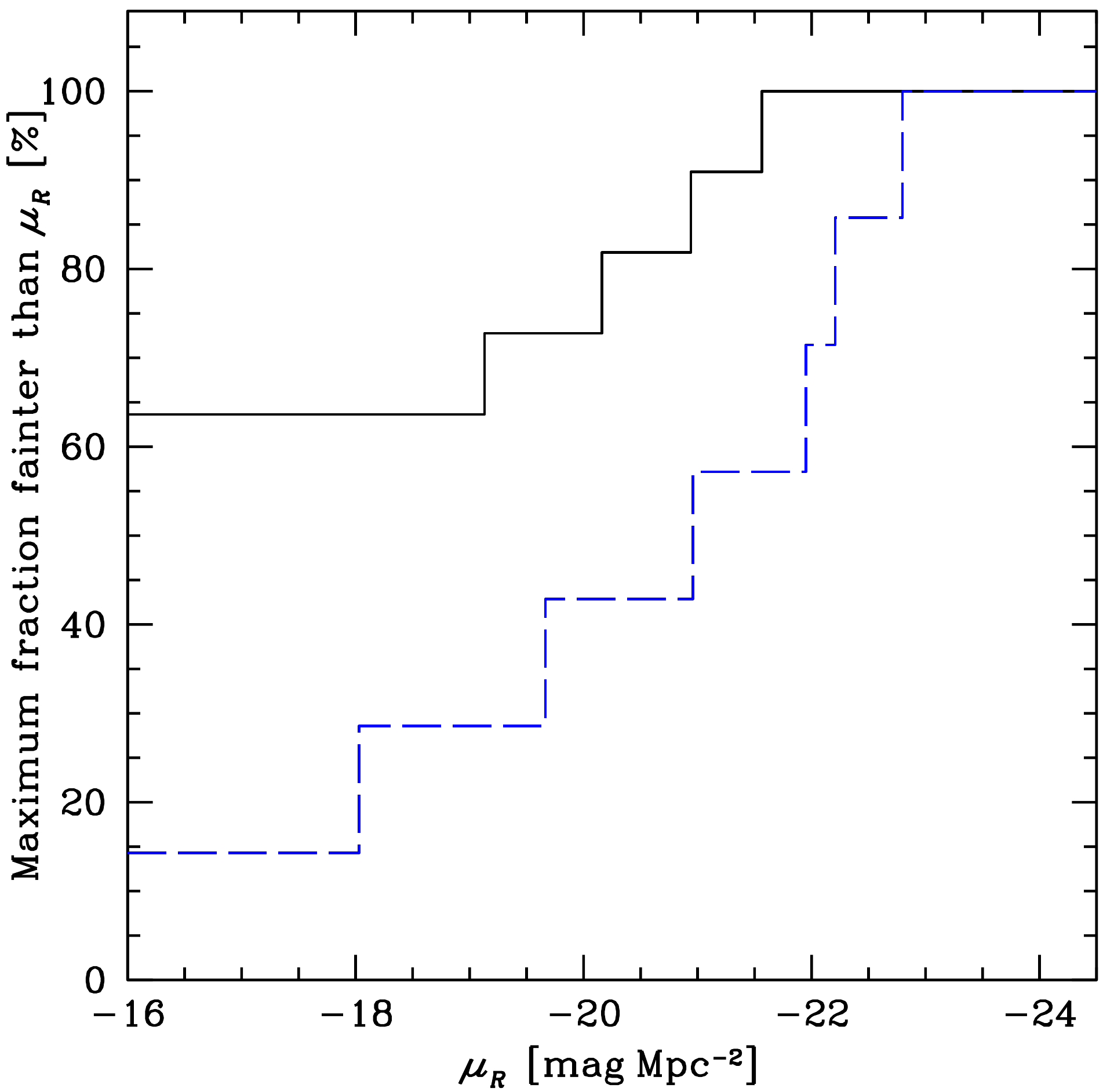}
	\caption{Cumulative fraction of absorbers originating in
          environments with no more than $N_{\rm
            gal}$ galaxies within a radius of $\rho=500$ kpc ({\it left panel}) and
          with mean surface brightness (averaged over the area of 500
          kpc radius) fainter than rest-frame $R$-band magnitude
          $\mu_R$ per square Mpc ({\it right panel}).  While a larger
          fraction of \lya-only absorbers ($>50$\% versus 16\% for
          O\,VI absorbers) arise in environments where no luminous
          galaxies are found within a radius of $\rho=500$ kpc, no
          statistically significant distinction can be made between
          the cumulative fraction of galaxy surface density around
          O\,VI and \lya-only absorbers based on these small samples.
          Specifically, a Kolmogorov-Smirnov (KS) test on their
          cumulative distribution functions shows that the probability
          that O\,VI and \lya-only absorbers are drawn from a similar
          parent galaxy environment distributions is at most 20\% (galaxy density)
          and 8\% (surface brightness).}
	\label{figure:ensemble}
\end{figure*}

The exercise presented in \S\,\ref{section:radial} shows that the ensemble average of
the radial profiles of galaxy surface density and surface brightness
exhibit different characteristics at $\rho\lesssim 500$ kpc between
O\,VI and \lya-only absorbers.  Here we focus on the inner regions of
$\rho=500$ kpc radius around individual absorbers and examine whether there
is a difference in the variation of galaxy properties within each
subsample.  This exercise provides further insights into the immediate galactic
environments of individual absorbers.

The left panel of Figure \ref{figure:ensemble} shows the cumulative fraction of absorbers
originating in environments of no more than $N_{\rm gal}$ galaxies.
It shows that while 16\% of O\,VI absorbers occur in environments where no
galaxies are found within a radius of $\rho=500$ kpc, more than 50\% of
\lya-only absorbers occur in such ``voids''.  However, the samples are
small and a Kolmogorov-Smirnov (KS) test shows that the probability
that both types of absorbers are drawn from a similar underlying
environment distribution is 20\%.

Next, we examine the mean surface density of star light averaged over
the area within 500 kpc radius in units of rest-frame $R$-band
magnitude per square Mpc.  This quantity allows us to account for
missing galaxies that are too faint to be detected in our survey by
including the survey limit in our analysis.  Including non-detections,
the right panel of Figure \ref{figure:ensemble} shows the maximum possible cumulative
fraction of absorbers arising in environments with mean galaxy surface
brightness fainter than the designated value $\mu_R$.  Similar to the
surface density plot in the left panel, we find that as much as 60\%
of \lya-only absorbers originate in environments with galaxy surface
brightness fainter than than $\mu_R\approx -19 \, {\rm mag \, Mpc^{-2}}$,
 while no more than 35\% of O\,VI absorbers are found in such low 
surface brightness environments.  A KS test shows that there is at 
most 8\% probability that O\,VI and \lya-only absorbers trace the same
underlying galaxy population.

\section{Discussion and Conclusions}
\label{section:discussion}
The high level of completeness achieved at faint magnitudes by our
survey has allowed us to probe the galaxy populations to low
luminosities and obtain better understanding of the nature of the
absorbing systems.  We have shown that O\,VI absorbers previously
attributed to a gaseous medium connecting massive galaxy groups are,
in fact, more closely associated with less massive groups containing
at least one dwarf, emission-line galaxy. In total, four of seven
known O\,VI absorbers along the PKS\,0405$-$123 sightline reside in
galaxy ``groups'' that contain at least one star-forming member.  Two
of the seven are found nearby to an emission-line galaxy and only one
O\,VI absorber is found in a galaxy void.  Therefore, we conclude that
O\,VI absorbers primarily trace gas-rich environments as indicated by
the presence of at least one low-mass, emission-line galaxy seen in
our survey. However, the presence of O$^{5+}$ ions could either be
the result of starburst driven outflows or due to stripped material
from galaxy interactions that also trigger star formation.

This is in stark contrast to lower ionization transitions such as
Mg\,II that are found to reside primarily in the halos of isolated
galaxies both with and without star formation \citep{Steidel:1997}.
One exception is those ultra-strong Mg\,II absorbers with rest-frame
absorption equivalent width $W_r(2796)>3$ \AA.  Three of these
ultra-strong Mg\,II systems have been targeted for follow-up galaxy
surveys and all three are found in groups containing multiple
super-$L_*$ galaxies \citep[e.g.][]{Whiting:2006, Nestor:2011, Gauthier:2013},
 with an inferred group halo mass of $M_h \sim 10^{13}\,{\rm
  M}_\odot$.  The galaxy ``groups'' found around O\,VI absorbers in
our survey contain primarily low-luminosity (and presumably low-mass)
galaxies \citep[see also][]{Mulchaey:2009}, and are therefore not as
massive as those found near ultra-strong Mg\,II absorbers.  A detailed
comparison of the dynamics between galaxies and absorbing gas may lead
to further insights into the physical origin of these O\,VI absorbers.

In addition, our analysis reveals a clear distinction in the radial
profiles of mean galaxy surface density and surface brightness around
different absorbers.  Specifically, O\,VI absorbers are found to
reside in galaxy overdensities with significantly higher mean galaxy
surface density and surface brightness at $\rho\lesssim 500$ kpc,
while only a mild increase in galaxy surface brightness is seen at
small $\rho$ around strong \lya-only absorbers.

On the other hand, \cite{Chen:2009} showed that both strong \lya\
absorbers of $\log\,N({\rm H\,I})\ge 14$ and \ovi\ absorbers exhibit a
comparable clustering amplitude as emission-line dominated galaxies.
The apparent discrepancy between the finding of this paper and those
of \cite{Chen:2009} may be explained by the intrinsic difference in
the sample definition.  Our current study is based on galaxies and
absorbers found in a single field, and therefore limited by the small
sample size.  In particular, we have defined a controlled \lya-only
sample that includes \lya\ absorbers that are as weak as $\log\,N({\rm
  H\,I})\approx13.6$ in order to have a sufficiently large sample for
a statistical analysis.  In contrast, the O\,VI absorbers in the
comparison sample all have associated \lya\ with $\log\,N({\rm
  H\,I})>14$.  It is therefore unclear whether the observed
distinction between the galactic environments of O\,VI and \lya-only
absorbers represents a fundamental difference between metal and H\,I
absorbers, or between high and low H\,I column density clouds.  We
expect that such uncertainty in the interpretation of the observations
will be resolved with a larger sample of galaxy and absorber data from
different fields complemented by multi-band imaging data.  Together,
these data will help constrain galaxy star-formation rates and reveal
the presence of any faint tidal features \citep[e.g.][]{Chen:2009}.

Lastly, an intriguing outcome of our deep galaxy survey is the
discovery of an O\,VI absorber that is likely to reside in a ``void''.
This O\,VI absorption system at $z\approx0.183$, which exhibits two
strong components separated by 70 \kms, does not have other ionic
transitions detected to sensitive limits (Table 3; see also Prochaska
\etal\ 2004).  Our galaxy survey data rule out the presence of any
galaxies of $L>0.04\,L_*$ at $\rho<250$ kpc or the presence of any
galaxies of $L>0.3\,L_*$ at $\rho<1$ Mpc.  The lack of additional
ionic transitions associated with this absorber suggests that the gas
may be hot and collisionally ionized.  Also, the
lack of galaxies found in the vicinity suggests that this absorber resides in an apparent
void, similar to an O VI absorber at z=0.06807 for which \cite{Tripp:2006} ruled out the
presence of galaxies of $L > 0.04\,L_*$ at $\rho < 200$ kpc.
While a likely explanation for the origin of the gas is the warm-hot
intergalactic medium, it is unclear whether the relatively narrow line
width is consistent with a diffuse intergalactic origin.

\section*{Acknowledgements}
It is a pleasure to thank Jean-Ren\'e Gauthier and the anonymous
referee for helpful comments that significantly improved the
manuscript.  We thank Edward Villanueva and Daniel Kelson for their
help with the COSMOS reduction pipeline and Joe Hennawi for his aid
with the Low-Redux reduction pipeline. We made extensive use of the
XIDL library provided by Jason Prochaska. We are grateful for the
support provided by the staff at the Las Campanas Observatory and the
Apache Point Observatory.  SDJ acknowledges funding from a National
Science Foundation Graduate Research Fellowship and a fellowship from
the Illinois Space Grant Consortium. This research has made use of the
NASA Astrophysics Data System and the NASA/IPAC Extragalactic
Database (NED) which is operated by the Jet Propulsion Laboratory,
California Institute of Technology, under contract with the National
Aeronautics and Space Administration. This paper contains data
obtained with the 6.5-m Magellan Telescopes located at Las Campanas
Observatory, Chile and data obtained with the Apache Point Observatory
3.5-meter telescope, which is owned and operated by the Astrophysical
Research Consortium.

\bibliographystyle{mn2e}
\bibliography{biblio}

\begin{thebibliography}{}

\bibitem[\protect\citeauthoryear{{Blanton}, {Hogg}, {Bahcall}, {Brinkmann},
  {Britton}, {Connolly}, {Csabai}, {Fukugita}, {Loveday}, {Meiksin}, {Munn},
  {Nichol}, {Okamura}, {Quinn}, {Schneider}, {Shimasaku}, {Strauss} \& {...
  Weinberg}}{{Blanton} et~al.}{2003}]{Blanton:2003}
{Blanton} M.~R.,  {Hogg} D.~W.,  {Bahcall} N.~A.,  {Brinkmann} J.,  {Britton}
  M.,  {Connolly} A.~J.,  {Csabai} I.,  {Fukugita} M.,  {Loveday} J.,
  {Meiksin} A.,  {Munn} J.~A.,  {Nichol} R.~C.,  {Okamura} S.,  {Quinn} T.,
  {Schneider} D.~P.,  {Shimasaku} K.,  {Strauss} M.~A.,    {... Weinberg}
  D.~H.,  2003, ApJ, 592, 819

\bibitem[\protect\citeauthoryear{{Burles} \& {Tytler}}{{Burles} \&
  {Tytler}}{1996}]{Burles:1996}
{Burles} S.,  {Tytler} D.,  1996, ApJ, 460, 584

\bibitem[\protect\citeauthoryear{{Carswell}, {Webb}, {Baldwin} \&
  {Atwood}}{{Carswell} et~al.}{1987}]{Carswell:1987}
{Carswell} R.~F.,  {Webb} J.~K.,  {Baldwin} J.~A.,    {Atwood} B.,  1987, ApJ,
  319, 709

\bibitem[\protect\citeauthoryear{{Cen}}{{Cen}}{2012}]{Cen:2012}
{Cen} R.,  2012, ApJ, 753, 17

\bibitem[\protect\citeauthoryear{{Cen} \& {Ostriker}}{{Cen} \&
  {Ostriker}}{1999}]{Cen:1999}
{Cen} R.,  {Ostriker} J.~P.,  1999, ApJ, 514, 1

\bibitem[\protect\citeauthoryear{{Chen} \& {Mulchaey}}{{Chen} \&
  {Mulchaey}}{2009}]{Chen:2009}
{Chen} H.-W.,  {Mulchaey} J.~S.,  2009, ApJ, 701, 1219

\bibitem[\protect\citeauthoryear{{Chen} \& {Prochaska}}{{Chen} \&
  {Prochaska}}{2000}]{Chen:2000}
{Chen} H.-W.,  {Prochaska} J.~X.,  2000, ApJL, 543, L9

\bibitem[\protect\citeauthoryear{{Chen}, {Prochaska}, {Weiner}, {Mulchaey} \&
  {Williger}}{{Chen} et~al.}{2005}]{Chen:2005}
{Chen} H.-W.,  {Prochaska} J.~X.,  {Weiner} B.~J.,  {Mulchaey} J.~S.,
  {Williger} G.~M.,  2005, ApJL, 629, L25

\bibitem[\protect\citeauthoryear{{Coleman}, {Wu} \& {Weedman}}{{Coleman}
  et~al.}{1980}]{Coleman:1980}
{Coleman} G.~D.,  {Wu} C.-C.,    {Weedman} D.~W.,  1980, ApJS, 43, 393

\bibitem[\protect\citeauthoryear{{Danforth} \& {Shull}}{{Danforth} \&
  {Shull}}{2008}]{Danforth:2008}
{Danforth} C.~W.,  {Shull} J.~M.,  2008, ApJ, 679, 194

\bibitem[\protect\citeauthoryear{{Dressler}, {Bigelow}, {Hare}, {Sutin},
  {Thompson}, {Burley}, {Epps}, {Oemler}, {Bagish}, {Birk}, {Clardy},
  {Gunnels}, {Kelson}, {Shectman} \& {Osip}}{{Dressler}
  et~al.}{2011}]{Dressler:2011}
{Dressler} A.,  {Bigelow} B.,  {Hare} T.,  {Sutin} B.,  {Thompson} I.,
  {Burley} G.,  {Epps} H.,  {Oemler} A.,  {Bagish} A.,  {Birk} C.,  {Clardy}
  K.,  {Gunnels} S.,  {Kelson} D.,  {Shectman} S.,    {Osip} D.,  2011, PASP,
  123, 288

\bibitem[\protect\citeauthoryear{{Ellingson} \& {Yee}}{{Ellingson} \&
  {Yee}}{1994}]{Ellingson:1994}
{Ellingson} E.,  {Yee} H.~K.~C.,  1994, ApJS, 92, 33

\bibitem[\protect\citeauthoryear{{Gauthier}}{{Gauthier}}{2013}]{Gauthier:2013}
{Gauthier} J.-R.,  2013, MNRAS, 432, 1444

\bibitem[\protect\citeauthoryear{{Gehrels}}{{Gehrels}}{1986}]{Gehrels:1986}
{Gehrels} N.,  1986, ApJ, 303, 336

\bibitem[\protect\citeauthoryear{{Ghavamian}, {Aloisi}, {Lennon}, {Hartig},
  {Kriss}, {Oliveira}, {Massa}, {Keyes}, {Proffitt}, {Delker} \&
  {Osterman}}{{Ghavamian} et~al.}{2009}]{Ghavamian:2009}
{Ghavamian} P.,  {Aloisi} A.,  {Lennon} D.,  {Hartig} G.,  {Kriss} G.~A.,
  {Oliveira} C.,  {Massa} D.,  {Keyes} T.,  {Proffitt} C.,  {Delker} T.,
  {Osterman} S.,  2009, Technical report, {Preliminary Characterization of the
  Post- Launch Line Spread Function of COS}

\bibitem[\protect\citeauthoryear{{Green}, {Froning}, {Osterman}, {Ebbets},
  {Heap}, {Leitherer}, {Linsky}, {Savage}, {Sembach}, {Shull}, {Siegmund},
  {Snow}, {Spencer}, {Stern}, {Stocke}, {Welsh}, {B{\'e}land}, {Burgh} \&
  {Danforth}}{{Green} et~al.}{2012}]{Green:2012}
{Green} J.~C.,  {Froning} C.~S.,  {Osterman} S.,  {Ebbets} D.,  {Heap} S.~H.,
  {Leitherer} C.,  {Linsky} J.~L.,  {Savage} B.~D.,  {Sembach} K.,  {Shull}
  J.~M.,  {Siegmund} O.~H.~W.,  {Snow} T.~P.,  {Spencer} J.,  {Stern} S.~A.,
  {Stocke} J.,  {Welsh} B.,  {B{\'e}land} S.,  {Burgh} E.~B.,    {Danforth} C.,
   2012, ApJ, 744, 60

\bibitem[\protect\citeauthoryear{{Howk}, {Ribaudo}, {Lehner}, {Prochaska} \&
  {Chen}}{{Howk} et~al.}{2009}]{Howk:2009}
{Howk} J.~C.,  {Ribaudo} J.~S.,  {Lehner} N.,  {Prochaska} J.~X.,    {Chen}
  H.-W.,  2009, MNRAS, 396, 1875

\bibitem[\protect\citeauthoryear{{Kang}, {Ryu}, {Cen} \& {Song}}{{Kang}
  et~al.}{2005}]{Kang:2005}
{Kang} H.,  {Ryu} D.,  {Cen} R.,    {Song} D.,  2005, ApJ, 620, 21

\bibitem[\protect\citeauthoryear{{Lehner}, {Savage}, {Richter}, {Sembach},
  {Tripp} \& {Wakker}}{{Lehner} et~al.}{2007}]{Lehner:2007}
{Lehner} N.,  {Savage} B.~D.,  {Richter} P.,  {Sembach} K.~R.,  {Tripp} T.~M.,
    {Wakker} B.~P.,  2007, ApJ, 658, 680

\bibitem[\protect\citeauthoryear{{Moos}, {Cash} \& {\etal}}{{Moos}
  et~al.}{2000}]{Moos:2000}
{Moos} H.~W.,  {Cash} W.~C.,    {\etal} 2000, ApJL, 538, L1

\bibitem[\protect\citeauthoryear{{Mulchaey} \& {Chen}}{{Mulchaey} \&
  {Chen}}{2009}]{Mulchaey:2009}
{Mulchaey} J.~S.,  {Chen} H.-W.,  2009, ApJL, 698, L46

\bibitem[\protect\citeauthoryear{{Mulchaey}, {Davis}, {Mushotzky} \&
  {Burstein}}{{Mulchaey} et~al.}{1996}]{Mulchaey:1996}
{Mulchaey} J.~S.,  {Davis} D.~S.,  {Mushotzky} R.~F.,    {Burstein} D.,  1996,
  ApJ, 456, 80

\bibitem[\protect\citeauthoryear{{Narayanan}, {Savage}, {Wakker}, {Danforth},
  {Yao}, {Keeney}, {Shull}, {Sembach}, {Froning} \& {Green}}{{Narayanan}
  et~al.}{2011}]{Narayanan:2011}
{Narayanan} A.,  {Savage} B.~D.,  {Wakker} B.~P.,  {Danforth} C.~W.,  {Yao} Y.,
   {Keeney} B.~A.,  {Shull} J.~M.,  {Sembach} K.~R.,  {Froning} C.~S.,
  {Green} J.~C.,  2011, ApJ, 730, 15

\bibitem[\protect\citeauthoryear{{Nestor}, {Johnson}, {Wild}, {M{\'e}nard},
  {Turnshek}, {Rao} \& {Pettini}}{{Nestor} et~al.}{2011}]{Nestor:2011}
{Nestor} D.~B.,  {Johnson} B.~D.,  {Wild} V.,  {M{\'e}nard} B.,  {Turnshek}
  D.~A.,  {Rao} S.,    {Pettini} M.,  2011, MNRAS, 412, 1559

\bibitem[\protect\citeauthoryear{{Oppenheimer} \& {Dav{\'e}}}{{Oppenheimer} \&
  {Dav{\'e}}}{2006}]{Oppenheimer:2006}
{Oppenheimer} B.~D.,  {Dav{\'e}} R.,  2006, MNRAS, 373, 1265

\bibitem[\protect\citeauthoryear{{Oppenheimer} \& {Dav{\'e}}}{{Oppenheimer} \&
  {Dav{\'e}}}{2009}]{Oppenheimer:2009}
{Oppenheimer} B.~D.,  {Dav{\'e}} R.,  2009, MNRAS, 395, 1875

\bibitem[\protect\citeauthoryear{{Oppenheimer}, {Dav{\'e}}, {Katz}, {Kollmeier}
  \& {Weinberg}}{{Oppenheimer} et~al.}{2012}]{Oppenheimer:2012}
{Oppenheimer} B.~D.,  {Dav{\'e}} R.,  {Katz} N.,  {Kollmeier} J.~A.,
  {Weinberg} D.~H.,  2012, MNRAS, 420, 829

\bibitem[\protect\citeauthoryear{{Prochaska}, {Chen}, {Howk}, {Weiner} \&
  {Mulchaey}}{{Prochaska} et~al.}{2004}]{Prochaska:2004}
{Prochaska} J.~X.,  {Chen} H.-W.,  {Howk} J.~C.,  {Weiner} B.~J.,    {Mulchaey}
  J.,  2004, ApJ, 617, 718

\bibitem[\protect\citeauthoryear{{Prochaska}, {Weiner}, {Chen}, {Cooksey} \&
  {Mulchaey}}{{Prochaska} et~al.}{2011}]{Prochaska:2011survey}
{Prochaska} J.~X.,  {Weiner} B.,  {Chen} H.-W.,  {Cooksey} K.~L.,    {Mulchaey}
  J.~S.,  2011, ApJS, 193, 28

\bibitem[\protect\citeauthoryear{{Prochaska}, {Weiner}, {Chen} \&
  {Mulchaey}}{{Prochaska} et~al.}{2006}]{Prochaska:2006}
{Prochaska} J.~X.,  {Weiner} B.~J.,  {Chen} H.-W.,    {Mulchaey} J.~S.,  2006,
  ApJ, 643, 680

\bibitem[\protect\citeauthoryear{{Richter}, {Savage}, {Tripp} \&
  {Sembach}}{{Richter} et~al.}{2004}]{Richter:2004}
{Richter} P.,  {Savage} B.~D.,  {Tripp} T.~M.,    {Sembach} K.~R.,  2004, ApJS,
  153, 165

\bibitem[\protect\citeauthoryear{{Savage}, {Narayanan}, {Wakker}, {Stocke},
  {Keeney}, {Shull}, {Sembach}, {Yao} \& {Green}}{{Savage}
  et~al.}{2010}]{Savage:2010}
{Savage} B.~D.,  {Narayanan} A.,  {Wakker} B.~P.,  {Stocke} J.~T.,  {Keeney}
  B.~A.,  {Shull} J.~M.,  {Sembach} K.~R.,  {Yao} Y.,    {Green} J.~C.,  2010,
  ApJ, 719, 1526

\bibitem[\protect\citeauthoryear{{Smith}, {Hallman}, {Shull} \&
  {O'Shea}}{{Smith} et~al.}{2011}]{Smith:2011}
{Smith} B.~D.,  {Hallman} E.~J.,  {Shull} J.~M.,    {O'Shea} B.~W.,  2011, ApJ,
  731, 6

\bibitem[\protect\citeauthoryear{{Spinrad}, {Filippenko}, {Yee}, {Ellingson},
  {Blades}, {Bahcall}, {Jannuzi}, {Bechtold} \& {Dobrzycki}}{{Spinrad}
  et~al.}{1993}]{Spinrad:1993}
{Spinrad} H.,  {Filippenko} A.~V.,  {Yee} H.~K.,  {Ellingson} E.,  {Blades}
  J.~C.,  {Bahcall} J.~N.,  {Jannuzi} B.~T.,  {Bechtold} J.,    {Dobrzycki} A.,
   1993, AJ, 106, 1

\bibitem[\protect\citeauthoryear{{Steidel}, {Dickinson}, {Meyer}, {Adelberger}
  \& {Sembach}}{{Steidel} et~al.}{1997}]{Steidel:1997}
{Steidel} C.~C.,  {Dickinson} M.,  {Meyer} D.~M.,  {Adelberger} K.~L.,
  {Sembach} K.~R.,  1997, ApJ, 480, 568

\bibitem[\protect\citeauthoryear{{Stinson}, {Brook}, {Prochaska}, {Hennawi},
  {Shen}, {Wadsley}, {Pontzen}, {Couchman}, {Quinn}, {Macci{\`o}} \&
  {Gibson}}{{Stinson} et~al.}{2012}]{Stinson:2012}
{Stinson} G.~S.,  {Brook} C.,  {Prochaska} J.~X.,  {Hennawi} J.,  {Shen} S.,
  {Wadsley} J.,  {Pontzen} A.,  {Couchman} H.~M.~P.,  {Quinn} T.,  {Macci{\`o}}
  A.~V.,    {Gibson} B.~K.,  2012, MNRAS, 425, 1270

\bibitem[\protect\citeauthoryear{{Stocke}, {Keeney}, {Danforth}, {Shull},
  {Froning}, {Green}, {Penton} \& {Savage}}{{Stocke}
  et~al.}{2013}]{Stocke:2013}
{Stocke} J.~T.,  {Keeney} B.~A.,  {Danforth} C.~W.,  {Shull} J.~M.,  {Froning}
  C.~S.,  {Green} J.~C.,  {Penton} S.~V.,    {Savage} B.~D.,  2013, ApJ, 763,
  148

\bibitem[\protect\citeauthoryear{{Stocke}, {Penton}, {Danforth}, {Shull},
  {Tumlinson} \& {McLin}}{{Stocke} et~al.}{2006}]{Stocke:2006}
{Stocke} J.~T.,  {Penton} S.~V.,  {Danforth} C.~W.,  {Shull} J.~M.,
  {Tumlinson} J.,    {McLin} K.~M.,  2006, ApJ, 641, 217

\bibitem[\protect\citeauthoryear{{Tepper-Garc{\'{\i}}a}, {Richter}, {Schaye},
  {Booth}, {Dalla Vecchia}, {Theuns} \& {Wiersma}}{{Tepper-Garc{\'{\i}}a}
  et~al.}{2011}]{Tepper:2011}
{Tepper-Garc{\'{\i}}a} T.,  {Richter} P.,  {Schaye} J.,  {Booth} C.~M.,  {Dalla
  Vecchia} C.,  {Theuns} T.,    {Wiersma} R.~P.~C.,  2011, MNRAS, 413, 190

\bibitem[\protect\citeauthoryear{{Thom} \& {Chen}}{{Thom} \&
  {Chen}}{2008}]{Thom:2008a}
{Thom} C.,  {Chen} H.-W.,  2008, ApJ, 683, 22

\bibitem[\protect\citeauthoryear{{Tilton}, {Danforth}, {Shull}, {Ross} \&
  {Ross}}{{Tilton} et~al.}{2012}]{Tilton:2012}
{Tilton} E.~M.,  {Danforth} C.~W.,  {Shull} J.~M.,  {Ross} T.~L.,    {Ross}
  T.~L.,  2012, ApJ, 759, 112

\bibitem[\protect\citeauthoryear{{Tripp}, {Aracil}, {Bowen} \&
  {Jenkins}}{{Tripp} et~al.}{2006}]{Tripp:2006}
{Tripp} T.~M.,  {Aracil} B.,  {Bowen} D.~V.,    {Jenkins} E.~B.,  2006, ApJL,
  643, L77

\bibitem[\protect\citeauthoryear{{Tripp} \& {Savage}}{{Tripp} \&
  {Savage}}{2000}]{TrippSavage:2000}
{Tripp} T.~M.,  {Savage} B.~D.,  2000, ApJ, 542, 42

\bibitem[\protect\citeauthoryear{{Tripp}, {Savage} \& {Jenkins}}{{Tripp}
  et~al.}{2000}]{Tripp:2000}
{Tripp} T.~M.,  {Savage} B.~D.,    {Jenkins} E.~B.,  2000, ApJL, 534, L1

\bibitem[\protect\citeauthoryear{{Tripp}, {Sembach}, {Bowen}, {Savage},
  {Jenkins}, {Lehner} \& {Richter}}{{Tripp} et~al.}{2008}]{Tripp:2008}
{Tripp} T.~M.,  {Sembach} K.~R.,  {Bowen} D.~V.,  {Savage} B.~D.,  {Jenkins}
  E.~B.,  {Lehner} N.,    {Richter} P.,  2008, ApJS, 177, 39

\bibitem[\protect\citeauthoryear{{Tumlinson}, {Thom}, {Werk}, {Prochaska},
  {Tripp}, {Weinberg}, {Peeples}, {O'Meara}, {Oppenheimer}, {Meiring}, {Katz},
  {Dav{\'e}}, {Ford} \& {Sembach}}{{Tumlinson} et~al.}{2011}]{Tumlinson:2011}
{Tumlinson} J.,  {Thom} C.,  {Werk} J.~K.,  {Prochaska} J.~X.,  {Tripp} T.~M.,
  {Weinberg} D.~H.,  {Peeples} M.~S.,  {O'Meara} J.~M.,  {Oppenheimer} B.~D.,
  {Meiring} J.~D.,  {Katz} N.~S.,  {Dav{\'e}} R.,  {Ford} A.~B.,    {Sembach}
  K.~R.,  2011, Science, 334, 948

\bibitem[\protect\citeauthoryear{{Wakker} \& {Savage}}{{Wakker} \&
  {Savage}}{2009}]{Wakker:2009}
{Wakker} B.~P.,  {Savage} B.~D.,  2009, ApJS, 182, 378

\bibitem[\protect\citeauthoryear{{Whiting}, {Webster} \& {Francis}}{{Whiting}
  et~al.}{2006}]{Whiting:2006}
{Whiting} M.~T.,  {Webster} R.~L.,    {Francis} P.~J.,  2006, MNRAS, 368, 341

\bibitem[\protect\citeauthoryear{{Williger}, {Heap}, {Weymann}, {Dav{\'e}},
  {Ellingson}, {Carswell}, {Tripp} \& {Jenkins}}{{Williger}
  et~al.}{2006}]{Williger:2006}
{Williger} G.~M.,  {Heap} S.~R.,  {Weymann} R.~J.,  {Dav{\'e}} R.,  {Ellingson}
  E.,  {Carswell} R.~F.,  {Tripp} T.~M.,    {Jenkins} E.~B.,  2006, ApJ, 636,
  631

\bibitem[\protect\citeauthoryear{{Woodgate, B. E., et al.}}{{Woodgate, B. E.,
  et al.}}{1998}]{Woodgate:1998}
{Woodgate, B. E., et al.} 1998, PASP, 110, 1183

\bibitem[\protect\citeauthoryear{{Yoon}, {Putman}, {Thom}, {Chen} \&
  {Bryan}}{{Yoon} et~al.}{2012}]{Yoon:2012}
{Yoon} J.~H.,  {Putman} M.~E.,  {Thom} C.,  {Chen} H.-W.,    {Bryan} G.~L.,
  2012, ApJ, 754, 84

\bibitem[\protect\citeauthoryear{{York}, {Adelman}, {Anderson} Jr. \&
  {\etal}}{{York} et~al.}{2000}]{York:2000}
{York} D.~G.,  {Adelman} J.,  {Anderson} Jr. J.~E.,    {\etal} 2000, AJ, 120,
  1579

\end{thebibliography}

\bsp
\label{lastpage}

\clearpage

\newpage

\begin{table*}
  \caption{Summary of spectroscopically identified galaxies at $\rho < 1$ Mpc and $|\Delta v| < 300$ \kms\, of O\,VI absorbers}
	\label{table:galaxies_OVI}
	\begin{tabular}{cccccccccccc}
	\hline
	\hline
	 & R.A. & Decl. & $\Delta \alpha$ & $\Delta \delta$ & $\Delta \theta$  & $\rho$ & R &  &  & $$ &  \\
	ID & (J2000) & (J2000) & ($''$) & ($''$) & ($''$)  & (kpc) & (mag) & $z_{\rm spec}$ & galaxy class$^{\rm a}$ & $L/L_*$ \\
\hline
\multicolumn{9}{l}{$z_{\rm sys} = 0.09180$ \,\,\,\, $\log N({\rm H\,I}) = 14.52$ \,\,\,\, $\log N({\rm O\,VI})=13.83$ \,\,\,\, $N_{\rm gal} = 5$}\\
\hline
$1835$ &04:07:49.4 & -12:12:16 & $14.2$ & $-39.3$ & $41.8$ & $71$ & $21.31 \pm0.10$ & $0.0923$ & E & $0.02$ \\
$2055$ &04:07:44.4 & -12:11:24 & $-59.1$ & $12.7$ & $60.4$ & $102$ & $21.30 \pm0.09$ & $0.0908$ & E & $0.02$ \\
$2080$ &04:07:43.2 & -12:11:48 & $-76.7$ & $-11.3$ & $77.5$ & $132$ & $21.27 \pm0.09$ & $0.0914$ & E & $0.02$ \\
$2212$ &04:07:40.2 & -12:13:44 & $-120.7$ & $-127.3$ & $175.4$ & $299$ & $21.94 \pm0.13$ & $0.0917$ & E & $0.01$ \\
$1698$ &04:07:52.6 & -12:15:49 & $61.1$ & $-252.3$ & $259.6$ & $439$ & $19.74 \pm0.07$ & $0.0908$ & E & $0.07$ \\
\hline
\multicolumn{9}{l}{$z_{\rm sys} = 0.09658$ \,\,\,\, $\log N({\rm H\,I}) = 14.65 $ \,\,\,\, $\log N({\rm O\,VI}) = 13.70$ \,\,\,\, $N_{\rm gal} = 5$}\\
\hline
$1457$ &04:07:58.1 & -12:12:24 & $141.8$ & $-47.3$ & $149.5$ & $267$ & $19.03 \pm0.07$ & $0.0966$ & E & $0.15$ \\
$1602$ &04:07:54.2 & -12:14:45 & $84.6$ & $-188.3$ & $206.5$ & $370$ & $19.01 \pm0.07$ & $0.0967$ & E & $0.15$ \\
$1601$ &04:07:54.2 & -12:14:50 & $84.6$ & $-193.3$ & $211.0$ & $379$ & $16.74 \pm0.06$ & $0.0969$ & E & $1.24$ \\
$2254$ &04:07:39.7 & -12:05:42 & $-128.0$ & $354.7$ & $377.1$ & $679$ & $18.74 \pm0.07$ & $0.0973$ & E & $0.20$ \\
$1659$ &04:07:54.4 & -12:03:07 & $87.5$ & $509.7$ & $517.1$ & $931$ & $17.99 \pm0.06$ & $0.0972$ & E & $0.39$ \\
\hline
\multicolumn{9}{l}{$z_{\rm sys} = 0.16711$ \,\,\,\, $\log N({\rm H\,I}) = 16.41$ \,\,\,\, $\log N({\rm O\,VI})= 14.77 $ \,\,\,\, $N_{\rm gal} = 2$}\\
\hline
$80006$ &04:07:48.3 & -12:11:03 & $-1.9$ & $33.7$ & $33.7$ & $96$ & $21.04 \pm0.00$ & $0.1669$ & E & $0.08$ \\
$1753$ &04:07:51.2 & -12:11:38 & $40.6$ & $-1.3$ & $40.6$ & $116$ & $17.43 \pm0.06$ & $0.1672$ & E & $2.13$ \\
\hline
\multicolumn{9}{l}{$z_{\rm sys} = 0.18259$ \,\,\,\, $\log N({\rm H\,I}) = 14.79$ \,\,\,\, $\log N({\rm O\,VI})= 14.12 $ \,\,\,\, $N_{\rm gal} = 0$}\\
\hline
\multicolumn{9}{l}{no galaxies of $L>0.04\,L_*$ at $\rho<250$ kpc and no galaxies of $L>0.3\,L_*$ at $\rho<1$ Mpc}\\
\hline
\multicolumn{9}{l}{$z_{\rm sys} = 0.29762$ \,\,\,\, $\log N({\rm H\,I}) = 14.00$ \,\,\,\, $\log N({\rm O\,VI}) = 13.61$ \,\,\,\, $N_{\rm gal} = 1$}\\
\hline
$1786$ &04:07:50.6 & -12:12:25 & $31.8$ & $-48.3$ & $57.9$ & $256$ & $19.34 \pm0.07$ & $0.2978$ & E & $1.38$ \\
\hline
\multicolumn{9}{l}{$z_{\rm sys} =  0.36078$ \,\,\,\, $\log N({\rm H\,I}) = 15.20$ \,\,\,\, $\log N({\rm O\,VI}) = 13.94$ \,\,\,\, $N_{\rm gal} = 2$}\\
\hline
$1967$ &04:07:45.9 & -12:11:09 & $-37.1$ & $27.7$ & $46.3$ & $233$ & $18.58 \pm0.07$ & $0.3614$ & A & $4.47$ \\
$1716$ &04:07:52.5 & -12:11:56 & $59.7$ & $-19.3$ & $62.7$ & $316$ & $23.01 \pm0.21$ & $0.3608$ & E & $0.08$ \\
\hline
\multicolumn{9}{l}{$z_{\rm sys} =  0.49510$ \,\,\,\, $\log N({\rm H\,I}) = 14.18$ \,\,\,\, $\log N({\rm O\,VI}) = 14.41$ \,\,\,\, $N_{\rm gal} = 1$}\\
\hline
$1862$ &04:07:49.1 & -12:11:21 & $9.8$ & $15.7$ & $18.5$ & $112$ & $22.63 \pm0.17$ & $0.4942$ & E & $0.25$ \\
\hline
\multicolumn{9}{l}{$^\mathrm{a}$ galaxy classification: E$\rightarrow$ emission-line dominated, A$\rightarrow$ absorption-line dominated.} \\
	\end{tabular}
\end{table*}

\begin{table*}
	\centering
  	\caption{Summary of spectroscopically identified galaxies at $\rho < 1$ Mpc and $|\Delta v| < 300$ \kms\, of \lya-only absorbers}
	\label{table:galaxies_HI}
	\begin{tabular}{cccccccccccc}
	\hline
	\hline
	 & R.A. & Decl. & $\Delta \alpha$ & $\Delta \delta$ & $\Delta \theta$  & $\rho$ & R &  &  & $$ &  \\
	ID & (J2000) & (J2000) & ($''$) & ($''$) & ($''$)  & (kpc) & (mag) & $z_{\rm spec}$ & galaxy class$^{\rm a}$ & $L/L_*$ \\
\hline
\multicolumn{9}{l}{$z_{\rm sys} =  0.08139$ \,\,\,\, $\log N({\rm H\,I}) = 13.79 $ \,\,\,\, $\log N({\rm O\,VI}) < 13.62 $ \,\,\,\, $N_{\rm gal} = 6$}\\
\hline
$2757$ &04:07:27.3 & -12:10:35 & $-309.8$ & $61.7$ & $315.9$ & $486$ & $17.70 \pm0.06$ & $0.0817$ & E & $0.35$ \\
$2804$ &04:07:26.5 & -12:11:12 & $-321.5$ & $24.7$ & $322.5$ & $496$ & $20.11 \pm0.07$ & $0.0817$ & A & $0.04$ \\
$2860$ &04:07:24.9 & -12:11:03 & $-345.0$ & $33.7$ & $346.6$ & $530$ & $20.40 \pm0.08$ & $0.0811$ & E & $0.03$ \\
$2867$ &04:07:24.5 & -12:10:41 & $-350.9$ & $55.7$ & $355.2$ & $545$ & $18.35 \pm0.07$ & $0.0815$ & A & $0.19$ \\
$1117$ &04:08:06.9 & -12:15:57 & $270.8$ & $-260.3$ & $375.6$ & $571$ & $18.81 \pm0.07$ & $0.0806$ & E & $0.12$ \\
$2110$ &04:07:43.2 & -12:02:49 & $-76.7$ & $527.7$ & $533.2$ & $812$ & $19.36 \pm0.07$ & $0.0807$ & E & $0.07$ \\
\hline
\multicolumn{9}{l}{$z_{\rm sys} =  0.13233 $ \,\,\,\, $\log N({\rm H\,I}) = 13.80$ \,\,\,\, $\log N({\rm O\,VI}) < 12.95 $ \,\,\,\, $N_{\rm gal} = 0$}\\
\hline
\multicolumn{9}{l}{no galaxies of $L>0.1\,L_*$ at $\rho<250$ kpc and no galaxies of $L>0.16\,L_*$ at $\rho<1$ Mpc}\\
\hline
\multicolumn{9}{l}{$z_{\rm sys} =  0.15304$ \,\,\,\, $\log N({\rm H\,I}) = 13.99$ \,\,\,\, $\log N({\rm O\,VI}) <  13.14$ \,\,\,\, $N_{\rm gal} = 2$}\\
\hline
$2034$ &04:07:44.0 & -12:12:09 & $-65.0$ & $-32.3$ & $72.6$ & $193$ & $18.24 \pm0.07$ & $0.1534$ & E & $0.84$ \\
$1334$ &04:08:01.6 & -12:08:36 & $193.1$ & $180.7$ & $264.4$ & $701$ & $21.75 \pm0.13$ & $0.1525$ & E & $0.03$ \\
\hline
\multicolumn{9}{l}{$z_{\rm sys} =  0.16121$ \,\,\,\, $\log N({\rm H\,I}) = 13.84$ \,\,\,\, $\log N({\rm O\,VI}) <  12.85$ \,\,\,\, $N_{\rm gal} = 3$}\\
\hline
$1503$ &04:07:57.4 & -12:07:41 & $131.5$ & $235.7$ & $269.9$ & $752$ & $18.19 \pm0.07$ & $0.1619$ & E & $0.99$ \\
$1504$ &04:07:57.5 & -12:07:34 & $133.0$ & $242.7$ & $276.7$ & $771$ & $21.45 \pm0.11$ & $0.1619$ & E & $0.05$ \\
$2784$ &04:07:26.6 & -12:12:52 & $-320.1$ & $-75.3$ & $328.8$ & $912$ & $19.08 \pm0.07$ & $0.1611$ & E & $0.43$ \\
\hline
\multicolumn{9}{l}{$z_{\rm sys} =  0.17876$ \,\,\,\, $\log N({\rm H\,I}) = 13.61$ \,\,\,\, $\log N({\rm O\,VI}) < 13.16$ \,\,\,\, $N_{\rm gal} = 2$}\\
\hline
$2282$ &04:07:38.4 & -12:14:41 & $-147.1$ & $-184.3$ & $235.8$ & $711$ & $20.46 \pm0.08$ & $0.1783$ & E & $0.15$ \\
$1345$ &04:08:01.6 & -12:07:51 & $193.1$ & $225.7$ & $297.0$ & $898$ & $19.25 \pm0.07$ & $0.1790$ & A & $0.46$ \\
\hline
\multicolumn{9}{l}{$z_{\rm sys} =  0.24554$ \,\,\,\, $\log N({\rm H\,I}) =13.82 $ \,\,\,\, $\log N({\rm O\,VI}) < 12.89$ \,\,\,\, $N_{\rm gal} = 0$}\\
\hline
\multicolumn{9}{l}{no galaxies of $L>0.07\,L_*$ at $\rho<250$ kpc and no galaxies of $L>0.5\,L_*$ at $\rho<1$ Mpc}\\
\hline
\multicolumn{9}{l}{$z_{\rm sys} =  0.33402$ \,\,\,\, $\log N({\rm H\,I}) = 13.82$ \,\,\,\, $\log N({\rm O\,VI}) < 13.12$ \,\,\,\, $N_{\rm gal} = 0$}\\
\hline
\multicolumn{9}{l}{no galaxies of $L>0.13\,L_*$ at $\rho<250$ kpc and no galaxies of $L>0.5\,L_*$ at $\rho<1$ Mpc}\\
\hline
\multicolumn{9}{l}{$z_{\rm sys} =  0.35099$ \,\,\,\, $\log N({\rm H\,I}) = 14.40$ \,\,\,\, $\log N({\rm O\,VI}) < 13.00$ \,\,\,\, $N_{\rm gal} = 3$}\\
\hline
$1939$ &04:07:46.7 & -12:12:17 & $-25.4$ & $-40.3$ & $47.7$ & $235$ & $20.18 \pm0.08$ & $0.3506$ & E & $0.95$ \\
$1787$ &04:07:50.9 & -12:10:41 & $36.2$ & $55.7$ & $66.4$ & $329$ & $21.48 \pm0.11$ & $0.3521$ & E & $0.29$ \\
$1652$ &04:07:53.8 & -12:13:58 & $78.7$ & $-141.3$ & $161.8$ & $802$ & $20.95 \pm0.09$ & $0.3517$ & E & $0.47$ \\
\hline
\multicolumn{9}{l}{$z_{\rm sys} =  0.40571$ \,\,\,\, $\log N({\rm H\,I}) = 14.98$ \,\,\,\, $\log N({\rm O\,VI}) < 13.28$ \,\,\,\, $N_{\rm gal} = 1$}\\
\hline
$80008$ &04:07:46.8 & -12:14:02 & $-23.9$ & $-145.3$ & $147.3$ & $800$ & $20.93 \pm0.00$ & $0.4065$ & A & $0.69$ \\
\hline
\multicolumn{9}{l}{$z_{\rm sys} =  0.40886$ \,\,\,\, $\log N({\rm H\,I}) = 14.44$ \,\,\,\, $\log N({\rm O\,VI}) < 13.28$ \,\,\,\, $N_{\rm gal} = 1$}\\
\hline
$1975$ &04:07:45.9 & -12:12:14 & $-37.1$ & $-37.3$ & $52.6$ & $287$ & $22.70 \pm0.21$ & $0.4100$ & E & $0.14$ \\
\hline
\multicolumn{9}{l}{$z_{\rm sys} =  0.53830$ \,\,\,\, $\log N({\rm H\,I}) = 14.22$ \,\,\,\, $\log N({\rm O\,VI}) < 13.26$ \,\,\,\, $N_{\rm gal} = 0$}\\
\hline
\multicolumn{9}{l}{no galaxies of $L>0.5\,L_*$ at $\rho<250$ kpc and no galaxies of $L>0.5\,L_*$ at $\rho<1$ Mpc}\\
\hline
\multicolumn{9}{l}{$^\mathrm{a}$ galaxy classification: E$\rightarrow$ emission-line dominated, A$\rightarrow$ absorption-line dominated.} \\
	\end{tabular}
\end{table*}

\end{document}